\documentstyle[12pt]{article}
\input{psfig}
\newcounter{chanum}
\newcounter{eqnnum1}
\renewcommand{\theequation}{\arabic{chanum}.\arabic{eqnnum1}}
\newcommand{\marge}[1]{\marginpar{}}  
\newcommand{\Sl}[1]{{}}           
\newcommand{\newsec}[1]{\addtocounter{chanum}{1}\setcounter{eqnnum1}{0}
				\section{#1}}
\newcommand{\beq}[1]{\addtocounter{eqnnum1}{1}\Sl{#1}\begin{equation}
                                  \if#1\empty\else\label{#1}\fi}
\newcommand{\eeq}{\end{equation}}
\newcommand{\beqa}[1]{\addtocounter{eqnnum1}{1}\Sl{#1}\begin{eqnarray}
                                  \if#1\empty\else\label{#1}\fi}
\newcommand{\eeqa}{\end{eqnarray}}

\newcommand{\Eq}[1]{(\ref{#1})}
\newcommand{\nm}{\nonumber\\}
\newcommand{\up}{$\uparrow$}
\newcommand{\down}{$\downarrow$}
\newcommand{\la}{\langle}
\newcommand{\ra}{\rangle}

\newcommand{\1}{1\mbox{$\!\!\!\!\!$}\mbox{\large$\bigcirc$}}
\newcommand{\2}{2\mbox{$\!\!\!\!\!$}\mbox{\large$\bigcirc$}}
\newcommand{\3}{3\mbox{$\!\!\!\!\!$}\mbox{\large$\bigcirc$}}
\newcommand{\4}{4\mbox{$\!\!\!\!\!$}\mbox{\large$\bigcirc$}}
\newcommand{\5}{5\mbox{$\!\!\!\!\!$}\mbox{\large$\bigcirc$}}
\newcommand{\6}{6\mbox{$\!\!\!\!\!$}\mbox{\large$\bigcirc$}}
\newcommand{\7}{7\mbox{$\!\!\!\!\!$}\mbox{\large$\bigcirc$}}
\newcommand{\8}{8\mbox{$\!\!\!\!\!$}\mbox{\large$\bigcirc$}}

\newcommand{\0}{ \mbox{$\!\!\!\!\!$}\mbox{\large$\bigcirc$}}
\newcommand{\first}{$\ast$\mbox{$\!\!\!\!\!$}\mbox{\large$\bigcirc$}}

\begin{document}

\title{Propagation and organization \\ 
       in lattice random media} 
\author{Patrick Grosfils \thanks{E-mail address: {\tt pgrosfi@ulb.ac.be}}
\ and Jean Pierre Boon \thanks{E-mail address: {\tt jpboon@ulb.ac.be}}\\
{\em Center for Nonlinear Phenomena and Complex Systems} \\
{\em Universit\'{e} Libre de Bruxelles, 1050 - Bruxelles, Belgium}  \\
{}\\
E. G. D. Cohen \thanks{E-mail address: {\tt egdc@rockvax.rockefeller.edu}}\\
{\em The Rockefeller University}\\
{\em New York, NY 10021} \\
and \\
L.A. Bunimovich\thanks{E-mail: {\tt bunimovh@math.gatech.edu}}\\
{\em School of Mathematics, Georgia Institute of Technology}\\
{\em Atlanta, GA 30332}}

\date{\today}

\maketitle

\begin{abstract}

We show that a signal can propagate in a particular direction through a 
model random medium regardless of the precise state of the medium. 
As a prototype, we consider a point particle moving on a one-dimensional 
lattice whose sites are occupied by scatterers with the following properties: 
(i) the state of each site is defined by its {\em spin} (up or down); 
(ii) the particle arriving at a site is scattered forward (backward) 
if the spin is up (down); 
(iii) the state of the site is modified by the passage of the particle, i.e.
the spin of the site where a scattering has taken place, flips 
($\uparrow \Leftrightarrow \downarrow $).
We consider one dimensional and triangular lattices, for which we give a 
microscopic description of the dynamics, prove the propagation of a particle 
through the scatterers, and compute analytically its statistical properties. 
In particular we prove that, in one dimension, the average 
propagation velocity is $\langle c(q) \rangle  = 1/(3-2q)$, with $q$ the 
probability that a site has a spin $\uparrow$, and, in the triangular lattice,
the average propagation velocity is independent of the scatterers distribution:
$\langle c \rangle = 1/8$. In both cases, the origin of the propagation is 
a blocking mechanism, restricting the motion of the particle in the direction 
opposite to the ultimate propagation direction, and there is a specific 
re-organization of the spins after the passage of the particle.
A detailed  mathematical analysis of this phenomenon is, to the best of our 
knowledge, presented here for the first time.

\end{abstract}

\newsec{Motivation}

Discrete systems with simple microscopic dynamics can exhibit peculiar
behavior showing some degree of complexity on a large scale. 
These systems have raised particular interest because they can be viewed as 
paradigms for complex phenomena such as growth processes, signal propagation 
in random media, spatio-temporal structuring in excitable media, or 
evolutionary dynamics \cite{conway}.
In particular, phenomena such as anomalous diffusion and oscillatory 
propagation have been reported by Cohen and co-workers \cite{cohenetal} who 
studied, numerically and theoretically, the motion of a particle in 
two-dimensional Lorentz Lattice Gases whose sites are occupied by 
scatterers which deflect the particle according to an {\em a priori} given 
rule. The most striking behavior in the particle dynamics is observed when 
the rule includes a feed-back of the particle on the substrate: the passage 
of the particle modifies the scattering property of the visited sites, e.g. 
if a particle arriving at a site is scattered say to the right (R), the state 
of the site is changed (R $\Rightarrow$ L) such that on its next visit the 
particle will be scattered to the left (``flipping scatterers''). 
This ``interaction'' between the particle and the state of the scatterer 
modifies considerably the dynamics as compared to the classical Lorentz 
Lattice Gas. For instance, lattices with 
flipping scatterers can yield oscillatory motion with overall propagation 
\cite{cohenetal}, while random fixed scatterers produce closed trajectories. 
Bunimovich and Troubetzkoy  proved a number of theorems about the boundedness 
or unboundedness of the trajectories of the particle for certain lattice 
models, and also discussed some generalizations~\cite{bunimov}.

In order to illustrate the propagative behavior, we show in Fig.1 
the case of a random Delaunay lattice \cite{wang}: here the particle
arriving at a node is deflected with the largest possible angle, either to 
the right or to the left, depending on the state of the scatterer (which
after the passage of the particle goes into the reverse state);  
whatever the initial configuration of the scatterers on the lattice, the 
particle will, after a few time steps, {\em always} enter a propagation 
phase (see the propagation {\em strip} shown in Fig.1). 

Here we present a detailed mathematical analysis of unbounded 
trajectories in one-dimensional and two-dimensional (triangular)
lattices.  In particular we show that for {\it{any}} initial 
configuration of the scatterers, the particle will always propagate in a 
certain direction with a given average velocity.  The direction of the 
propagation is determined by the initial velocity of the particle and 
the initial state of the scatterers in a small neighborhood of the 
initial position of the particle. On the triangular lattice 
the entire trajectory becomes quickly confined to a particular strip, which 
is bounded by two adjacent parallel lines of the lattice \footnote{This 
propagation phenomenon was first observed and analyzed on the triangular 
lattice by Kong and Cohen \cite{cohenetal}.}. The propagation is due to a 
{\it{blocking mechanism}} which prevents the particle from moving in a 
direction opposite to the propagation direction for more than a few steps;
thus the particle can visit any lattice site at most three times in one
dimension, and six times on the triangular lattice. The blocking
mechanism is due to an organization of the scatterers induced by their 
interaction with the moving particle, which itself is thereby forced to 
propagate indefinitely in a particular direction. While in one dimension 
the rearrangement of the scatterers after the passage of the particle is a 
simple mapping of the initial configuration, on the triangular lattice a
self-organization of the scatterers takes place along the propagation 
strip: the visited sites on one of the two boundary lines are
occupied by only one kind (R or L) of scatterers, while the other  
boundary line is occupied by only the other kind (L or R). We find 
another important distinction: the average propagation velocity 
(averaged over all possible initial distributions of the scatterers
with a fixed ratio of R and L scatterers) in one dimension, depends 
on the {\em a priori} probability that a site has a spin $\uparrow$, 
while in the two-dimensional (triangular) lattice, it is a constant 
(independently of the probability that a site has a L or R scatterer). 

We first consider the one dimensional lattice.
In section \ref{sec:1D_latt}, we define the 1-D system 
considered here, introduce the basic equations describing the 
microscopic dynamics, and prove propagation of the particle; 
section \ref{sec:stat_prop} is devoted to a detailed study of the 
statistical properties of the particle motion.
The triangular lattice is treated in section \ref{sect:tri_lattice} 
where particle propagation and its general and statistical properties 
are analyzed. All our analytical results are shown to be in agreement with 
our corresponding numerical simulation data obtained for one- and 
two-dimensional lattices.   
We conclude  with some general comments (section \ref{sec:concl_comm}).

\newsec{The 1-D lattice}
\label{sec:1D_latt}

We consider a regular one-dimensional lattice where a single particle
moves from one site to the next, with the following properties: (i) the 
state of each site is defined by its {\em spin} (up or down); (ii) the 
particle arriving at a site is scattered forward (backward) if the spin 
is up (down); (iii) the spin of the site where a scattering has taken place, 
flips ($\uparrow \Rightarrow \downarrow , \downarrow \Rightarrow \uparrow$)
\footnote{A generalization to more complicated transitions in the scattering
properties of the sites has been considered for cyclic cellular automata 
by Bunimovich and Troubetzkoy  \cite{bunimov} who proved a number of  
general theorems, without having to use explicitly the equations 
of motion for the particle and the equations for the dynamics of the 
scatterers.}.
Setting the distance between neighboring sites and the 
speed of the particle equal to unity, the particle moves from one lattice 
site to the next in one unit time step . The initial position and velocity 
of the particle are arbitrarily fixed, and the spins are arbitrarily 
distributed on the lattice with an {\em a priori} probability $q$ that 
a site has a spin $\uparrow$.

\subsection{Definitions}
\label{subsec:def}

\noindent (i) Define the position $R(t)$ of the particle at time $t$ as 
the location of the site $r$ where the particle resides at time $t$.\\

\noindent (ii) Define the velocity of the particle: $C(t)$, whose value is 
$+\,1$ (when the particle moves to the right) or $-\,1$ (when it moves 
to the left) \footnote{With the convention that the positive direction 
on the 1-D lattice is from left to right, we can omit vectorial notation 
for simplicity.}. \\ 

\noindent (iii) Define the spin (the orientation of the scatterer)
\footnote{In  previous publications \cite{cohenetal} the scatterers were 
indicated as to their right or left scattering properties (R or L);
for the 1-D case a representation of the scatterers by up or down spins 
is more convenient.}
at site $r$, at time $t$, as a Boolean variable: $\eta(r,t)$, with value 
$+\,1$ ({\em spin up}, which does not modify the velocity of the particle),
or  $-\,1$ ({\em spin down }, which reverses the velocity of the 
particle from $C\,=\,+\,1$ to $C\,=\,-\,1$ or {\em vice versa}).

\subsection{Basic equations}
\label{subsec:basic_eqs}

(1) At each time step the particle moves to a neighboring site according to 
its velocity, and its new position is given by the equation of motion:
\beq{a2.1}
R(t+1)\,=\,R(t)\,+\,C(t)\,.
\eeq

\noindent (2) The velocity of the particle at its new position is maintained 
or reversed, depending on  the state of the spin (the orientation of the 
scatterer):
\beq{a2.2}
C(t+1)\,=\,C(t)\,\eta(R(t+1), t)\,.
\eeq
In $\eta(R(t+1), t)$, $t$ refers to the time corresponding to the state of 
the spin {\em before} flipping; we shall denote by $t+$ the time just 
{\em after} the particle arrived at a site and made its spin  flip 
($\uparrow \Leftrightarrow \downarrow $).\\

\noindent (3) The spin of the site hit by the particle is reversed
(all other spins remain unchanged):
\beqa{a2.3}
\eta(r, t+1)&=&\eta(r, t)\,(1-\delta_{r,R(t+1)})\,-\,\eta(r, t)\,
\delta_{r,R(t+1)}\nm &=&\eta(r, t)\,(1\,-\,2\,\delta_{r,R(t+1)})\,;
\eeqa
so  
\beqa{a2.4}
\eta(R(t+1), t+1)\,=\,-\,\eta(R(t+1), t)\,. 
\eeqa

Equations \Eq{a2.1}, \Eq{a2.2}, and \Eq{a2.4} are the basic equations of 
the dynamics on the 1-D lattice.

\subsection{Propagative motion}
\label{subsec:propag}

We say that the particle propagates in one direction on the 
one-dimensional lattice, if it visits any lattice site not more than 
some fixed finite number of times. This definition implies that the 
trajectory of the particle tends to $+ \infty$ or $- \infty$, when 
the time goes to infinity.  Indeed since the  particle cannot
return to a site after it has visited it a finite number of
times, the particle must remain on a semi-line to the right 
(or to the left) of this site, which holds equally for the next
visited site. Consequently the position of the particle must 
approach $+ \infty$ or $- \infty$ when $t \rightarrow \infty$.

We now formulate the propagation theorem.\\

\noindent{\em Theorem 1} : 
A particle moving from site to site in a one-dimensional lattice 
fully occupied with flipping scatterers (spins), propagates in one 
direction, independently of the initial distribution of the spins on the 
lattice.
The propagation  direction depends upon the orientation of the spin 
at the origin of the particle motion $(r=0)$ and of that at the
site $r=+1$, if the initial particle velocity $C(t=0+) \equiv C(0+)= +1,$  
or alternatively of that at site $r=-1$, if $C(0+)=-1$.\\

\noindent{\em Proof} : Without loss of generality we assume that 
the initial velocity of the particle is positive, i.e., $C(0+)=1$. 
Then we have to consider two cases according to whether the spin of the
site at the origin ($r=0, t=0$) is either \up\ or \down\ (see Fig.2 for
illustration of the proof).

A. If at $t=0+$, $\eta (R(0),0+) = \eta (0,0+)=-1$, there are two 
possibilities:\\
(1) the spin at time 0 at the position $R(1)$ is \up\ , 
i.e. $\eta (R(1),0) = \eta (1,0) = 1$. Then at time $1+$, $C(1+) = 1$ 
and $\eta (R(1), 1+) = \eta (1,1+)=-1$;\\
(2) the spin at time 0 at the position $R(1)$ is \down\ , 
i.e., $\eta (R(1),0) = \eta (1,0) = -1$. At time $1+$, $C(1+) = -1$ 
and $\eta (1,1+) = +1$, while at time 2+ ,
$C(2+) = +1$ and $\eta (R(2), 2+) = \eta (0,2+) = 1$. Then at time 3+,
$C(3+) = 1$ and $\eta(R(3), 3+) = \eta (1,3+) = -1$.

Therefore in both cases (1) and (2), the system is in a state where the
particle is at $r = 1$, leaving the visited sites with the 
velocity $+1$  ($C(1+)=1$ in case (1) and $C(3+)=1$ in case (2)),
while the scatterer at this site ($r = 1$) is \down\ . Thus we retrieve 
the initial situation shifted by one lattice unit to the right, and 
consequently the same analysis can be repeated for the next site, and so on.

We call the state of the system when $C(t+) = 1$ and $\eta (R(t),
t+) = -1$, a left {\em {blocking pattern}}.  So the above analysis
shows that a left blocking pattern at the site $R(t)$ will be
shifted to a left blocking pattern at the next site to the right
of $R(t)$ at $R(t)+1$, in one time step in case (1) ($R(t+1) = R(t) + 1$), 
or in 3 time steps in case (2) ($R(t+3) = R(t) + 1$), leading, 
in both cases, to propagation of the particle to the right.

B. If at $t=0+$, $\eta(R(0),0+) = \eta (0,0+) = +1$, there are again 
two possibilities:\\
(1) the spin at time 0 at the position $R(1)$ is \up\ ,
i.e. $\eta (R(1),0) = \eta(1,0) = +1$. Then $C(1+) = 1$ and 
$\eta(R(1),1+) = \eta (1, 1+) = -1$, and the same blocking pattern 
obtains at $r=1$ at the time $t = 1+$, as in case A(1).\\
(2) the spin at time 0 at the position $R(1)$ is \down\ ,
i.e. $\eta (1,0) = -1$. Therefore $C(1+) = -1$ and $\eta(R(1), 1+) =  
\eta(1,1+) = 1$. Then at time 2+, $C(2+) = -1$ and 
$\eta (R(2), 2+) = \eta (0,2+) = -1$, and we obtain a situation opposite
to that in Case A(2), which implies that there is now a {\em right} 
blocking pattern leading to a propagation to the left.
Thus the same analysis is applicable here, if one changes positive 
velocities and position changes to negative ones and vice-versa. 

We conclude (see Fig.2) that there are three cases where the initial 
conditions for $C$ and $\eta$ produce a left blocking mechanism and a 
propagation to the right: 
\begin{center} 
A(1): $\;\; C(0+)=+1,\; \eta(0,0+)=-1,\; \eta(1,0)=+1,$  \\
A(2): $\;\; C(0+)=+1,\; \eta(0,0+)=-1,\; \eta(1,0)=-1,$  \\
B(1): $\;\; C(0+)=+1,\; \eta(0,0+)=+1,\; \eta(1,0)=+1,$  
\end{center}
\noindent and one initial condition: 
\begin{center} 
B(2): $\;\; C(0+)=+1,\; \eta(0,0+)=+1,\; \eta(1,0)=-1,$  
\end{center}
\noindent which  produces a right blocking mechanism and a propagation 
to the left .

Clearly the same analysis holds if the initial velocity of the particle is 
to the left: for $C(0+) = -1$, there are three cases where the propagation 
is to the left, and one where the propagation is to the right.  Propagation 
in the direction of the initial velocity occurs in three out of four cases.
Notice that by {\em fixing} the initial condition with only {\em one} 
spin: $\eta(R(0),0+)=-1$, i.e. the spin is set \down\ at the origin, 
the particle always propagates in the direction of its initial velocity.

Two corollaries follow from {\em Theorem 1}.\\
\noindent {\em Corollary 1} :  
A particle moving on a 1-D lattice fully occupied
with flipping scatterers visits any lattice site at most three times.\\
The proof follows immediately from {\em Theorem 1} as a consequence
of the blocking mechanism.\\
\noindent {\em Corollary 2} :  The propagation velocity $C_{\ell}$ between 
two sites separated by a distance $\ell$ containing $u$ spins \up\ ,
is $C_{\ell} = (3 - 2u/\ell)^{-1}$.\\
{\em Proof} : We first note that the time $t_{\ell}$ taken by the particle
to cover the distance $\ell$ equals $\ell$ plus twice the number of times 
the particle visits a site with spin \down\ , forcing the particle into a 
backward-forward motion. 
Thus $t_{\ell} = \ell + 2 (\ell-u) = 3 \ell - 2u$.  Since 
$C_{\ell} = \ell /t_{\ell}$, the corollary follows immediately.

Finally we note that after the passage of the particle through the lattice, 
all spins of the visited sites have been reversed 
($\uparrow \Leftrightarrow \downarrow$) with a shift of one lattice site 
opposite to the ultimate velocity of propagation. This re-organization of
the spins, which is a consequence of the blocking mechanism, is illustrated 
in Fig.3.

\newsec{Statistical properties}
\label{sec:stat_prop}

Statistical properties are obtained by considering average quantities 
$\langle \cdots \rangle$, where the average is taken over all initial spin 
configurations, with the initial {\em a priori} probability $q$ that a site 
is in the state \up .

\subsection{Average velocity}
\label{subsec:av_veloc}

From {\em Corollary 2}, it follows immediately that the average time 
$\la t_{\ell}\ra$ taken by the particle to cover a finite segment of length 
$\ell$ on the 1-D lattice  is:
$\la t_{\ell} \ra = 3 \ell - 2 \la u \ra = \ell (3-2q)$.  Then the average
propagation velocity $\la c(q) \ra \equiv c(q)$ is given by
\beq{a4.1}
c(q) = \ell/\la t_{\ell} \ra = 1/(3-2q),
\eeq
independently of the direction of propagation and regardless of the way the 
spins are organized. Examples are shown in Fig.4.

\subsection{Time evolution}
\label{subsec:time_evol}

The equation of motion of the particle can be written in terms of two 
Boolean variables: \\ 
(i) the occupation variable $n(r,t)$, which is equal to $1$ if the particle 
is at site $r$ at time $t$ {\em for the first time}; otherwise $n(r,t) = 0$;\\
(ii) the state of the spin at site $r$ at time $t$: $\xi_{\uparrow}(r,t) = 1$ 
if the spin is \up\ , and $\xi_{\downarrow}(r,t)= 1$ if the spin is \down\ , 
with $\Sigma_{j=\uparrow,\downarrow}\,\xi_{j}=1$.\footnote{Notice the
difference between $\eta = \pm 1$ in section \ref{subsec:basic_eqs} 
\cite{cohen} and $\xi_j = \{0,1\}$, which is the quantity used in practice for 
numerical simulations.} Then the equation for $n(r,t)$ reads
\beq{a4.2}
n(r,t) \,=\, \xi_\uparrow  (r-1,0)\, n(r-1,t-1)\,+\,
\xi_\downarrow (r-1,0)\, n(r-1,t-3)\,,
\eeq
which is a microscopic equation.  

We now define $P_1(r,t) = \la n(r,t) \ra$, the probability that the particle, 
starting at the origin, visits site $r$ at time $t$ {\it{for the first time}}.
In the following we shall use $f(r,t) \equiv P_1(r,t)$ for short, with the
property $\sum_t P_1(r,t) \equiv  \sum_t f(r,t) = 1, \forall~r~>~0$, which
follows from the fact that each site will have a first visit. We also have 
$\la \xi_{\uparrow}(r,0)\ra = q$ and $\la \xi_{\downarrow}(r,0) \ra = 1-q,
\forall~r$. Averaging the basic equation \Eq{a4.2} over all initial 
configurations and using the fact that $\la \xi n \ra$ on the r.h.s. 
of \Eq{a4.2} can be written as $\la \xi \ra \la n \ra$ then yields 
\beqa{a4.3a}
f(r+1,t+1) \,=\, q\, f(r,t)\,+\,(1-q)\, f(r,t-2)\,,
\eeqa
which shows a formal analogy with the equation for the biased random walk 
(BRW)\footnote{The BRW equation \cite{randomwalk} reads
$g(r+1,t+1) \,=\, h\, g(r,t)\,+\,(1-h)\, g(r+2,t)\,,\;\;\;h\,\neq 0,1\,,$
where $g(r,t)$ is the distribution of the random walker which, at each time 
step, moves to the right or to the left  with probability $h$ and $1-h$ 
respectively (for $h=1/2$, the equation describes the usual random walk). 
For a comparison with Eq.\Eq{a4.3a}, see also \cite{boon}.}. 
For convenience in the forthcoming development, we perform a shift of 
variable ($t \rightarrow t+3$) to rewrite Eq.\Eq{a4.3a} as 
\beq{a4.3}
f(r+1,t+3) \,=\, q\, f(r,t+2)\,+\,(1-q)\, f(r,t)\,.
\eeq
Considering that the particle, arriving from the left, is at $r=0$ for the 
first time at $t=0$ with unknown velocity, and that the spin at $r=-1$ is 
down, the initial conditions are given by
\beqa{a4.4}
 f(r,t=0)\,=\,\delta_{r,0}\,;\;\;
 f(r,t=1)\,=\,q\,\delta_{r,+1}\,;\;\;
 f(r,t=2)\,=\,q^2\,\delta_{r,+2}\,.
\eeqa
Introducing the space-Fourier transform $f_k(t)\,=\,{\cal F} \{f(r,t)\}$ and 
its discrete Laplace transform $\tilde{f}_k(s)\,=\,{\cal L} \{f_k(t)\}$, we 
note that
\beqa{a4.5}
{\cal L \, F} \{f(r+n,t+m)\}\,&=& 
\,e^{{\imath}kn}[e^{ms}\tilde{f}_k(s) - e^{ms}f_k(0)-e^{(m-1)s}f_k(1)- \nm
&& \cdots -\,e^{s}f_k(m-1)]\,.
\eeqa 


Equation \Eq{a4.3} is then Fourier-Laplace transformed to yield
\beqa{a4.6}
e^{{\imath}k}e^s \left[ e^s [ e^s (\tilde{f}_k(s) - f_k(0)) - f_k(1)] -
f_k(2) \right] &=&  \nm
 q\, e^s [ e^s (\tilde{f}_k(s) - f_k(0)) - f_k(1)]
&+& (1 - q)\, \tilde{f}_k(s)  \,, 
\eeqa
with 
\beqa{a4.7}
f_k(0)\,=\,1\,;\;\;\; f_k(1)\,=\,q\, e^{{\imath}k}\,;\;\;\;
f_k(2)\,=\,q^2\,e^{{2\imath}k}\,,  
\eeqa
which follow from the initial conditions \Eq{a4.4}. Setting 
$z=e^s$ and $\kappa=e^{{\imath}k}$, the solution to Eq.(3.6) reads
\beqa{a4.8}
\tilde{f}_{\kappa}(z)\,=\,z\,\frac{z^2 + q(\kappa - 1/\kappa) +
q^2 (\kappa^2 - 1)}{z^2 (z- q/\kappa) - (1 - q)/\kappa} \,. 
\eeqa

The function $\tilde{f}_{\kappa}(z)$ can be explicitly inverted
in time and space for $q=1$ and $q=0$; the results for 
$f(r,t)\equiv P_1(r,t)$ are given, as expected, in terms of
$\delta$-functions, $\delta (r,t)$ for $q=1$ and $\delta (r,t/3)$ for $q=0$.
When $q \neq 0$ or $1$, we consider the case $\kappa = 1$, i.e. the limit 
of long wavelengths ($k=0$), for which the poles of the above transform can 
be computed easily \footnote{The computation for $ \kappa \neq 1$ (with 
$q \neq 0$ or $1$) is quite involved and the explicit analytical results 
are not very instructive.}; we obtain
\beqa{a4.9}
\tilde{f}_{\kappa = 1}(z)\,=\,z^3\,[(z-z_0)(z-z_+)(z-z_-)]^{-1} \,, 
\eeqa
where
\beqa{a4.10}
z_0\,=\,1\,;\;\; 
z_{\pm}\,=\,(1-q)^{1/2}(\cos \varpi \mp \imath\,\sin \varpi)\,,
\eeqa
with
\beqa{a4.11}
\cos \varpi\,=\,-\frac{1}{2}\,(1-q)^{1/2}\,,\;\;
\sin \varpi\,=\,\frac{1}{2}\,(3+q)^{1/2}\,.
\eeqa
The function $\tilde{f}_{\kappa =1}(z)$ is then inverted (using residues)
to obtain the explicit expression for $f_{k =0}(t) = \Sigma_r f(r,t)$; after
some straightforward algebra, we find
\beqa{a4.12}
f_{k =0}(t)\,=\,\frac{1}{3-2q}\,[1+2(1-q)\,e^{-\zeta t}\,(\cos \varpi t +
\varphi \,\sin \varpi t)]\,,
\eeqa
with 
\beqa{a4.13}
\zeta = \frac{1}{2}\,\log \frac{1}{1-q}\,,\;\;  
\varpi = \arctan -\sqrt{\frac{3+q}{1-q}},\;\; 
\varphi = \frac{q}{(1-q)^{1/2}(3+q)^{1/2}}.
\eeqa

The case $q=0$ provides a trivial but illustrative example ; it is 
straightforwardly verified that the solution then reads 
\beqa{a4.14}
f_{k =0}^{q=0}(t)\,=\,\frac{1}{3}\,(1+2 \cos \frac{2\pi}{3}t)\,,
\eeqa
which is a repeated sequence of values ($1,0,0$) as can be inferred
from the dynamics of the particle when all spins are initially down. The
case $q=1$ (all spins initially up) yields the obvious result 
$f_{k =0}^{q=1}(t)\,=\,1$. 

The long-time behavior of $f_{k =0}(t)$ is easily obtained from the limit
$s \rightarrow 0$ of $\tilde{f}_{k=0}(s)$ by expanding $\tilde{f}_{k=0}(s)$
to lowest order in $s$
\beqa{a4.15}
\lim_{s \rightarrow 0} \tilde{f}_{k=0}(s)\,=\,\frac{1}{s}
\,\frac{1}{3-2q}\,,
\eeqa
which yields by Laplace inversion 
\beqa{a4.16}
\lim_{t \rightarrow \infty} {f}_{k=0}(t)\,=\,
\frac{1}{3-2q}\,.
\eeqa
This result, combined with \Eq{a4.1}, shows that 
\beqa{a4.17}
\lim_{t \rightarrow \infty} \sum_{r} P_1(r,t)\,=\,c(q)\,.
\eeqa
In Fig.5 we show that the above analytical results and the simulation 
data are in perfect agreement. An alternative derivation of \Eq{a4.17},
directly based on the analysis of the particle dynamics, is given in 
Appendix A.

\subsection{Space evolution}
\label{subsec:space_evol}

We now construct an explicit analytical expression in $r$ and $t$
for the probability
that the particle visits, for the first time, site $r$ at time $t$, i.e.
the function $P_1(r,t)$.  Consider, on the 1-D lattice, a segment with 
length $\ell$ (in lattice units) and with spin configuration 
$\{\eta_i\}=(\eta_0, \eta_1, \cdots ,\eta_l)$, where $\eta_i =\, \uparrow$
or $\downarrow$. Suppose that at time zero, the particle is at site $r=0$ 
with velocity $C(0+)=+1$,  and the spin is $\eta_0=\,\downarrow$. Then the 
time $\tau$ taken by the particle to travel the distance 
$\ell$ is $ \tau(\ell,\{\eta_i\}) = 1 + N_{\uparrow} + 3N_{\downarrow}$, 
where $N_{\uparrow}$ and $N_{\downarrow}$ are the numbers of sites with 
spin up and spin down, respectively, between $r=1$ to $r=\ell-1$ (since 
the spin $\eta_\ell$ is unimportant). 
So $N_{\downarrow} = \ell -1 - N_{\uparrow}$ and 
\beqa{a3.18}
\tau(\ell,\{\eta\}) = 3\ell - 2(1 + N_{\uparrow}).
\eeqa 
The probability for the particle to be for the first time at site $r$ at 
time $t$ for a given spin configuration is
\beqa{a4.19}
P_1(r,t;\{\eta_i\})=\delta_{t,\tau (r,\{\eta_i\})},
\eeqa 
so that
\beqa{a4.20}
P_1(r,t)=\sum_{\{\eta_i\}} \delta_{t,\tau (r,\{\eta_i\})} P(\{\eta_i\})\,,
\eeqa 
where $P(\{\eta_i\})$ is the probability of the spin configuration 
$\{\eta_i\}$
\beqa{a4.21}
 P(\{\eta_i\})=\left( \begin{array}{c} r-1 \\ N_{\uparrow} 
\end{array} \right) q^{N_{\uparrow}} (1-q)^{r-1-N_{\uparrow}}\,,
\eeqa
with $q$ the probability that a site has spin ${\uparrow}$. Furthermore, 
using \Eq{a3.18} for $\ell = r$, we have
\beqa{a4.22}
P_1(r,t)&=&\sum_{N_{\uparrow}=0}^{r-1} \delta_{t,3r-2(1 + N_{\uparrow})}
\left( \begin{array}{c} r-1 \\ N_{\uparrow} \end{array} \right)
q^{N_{\uparrow}}\, (1-q)^{r-1-N_{\uparrow}} \nm   
&=&\left( \begin{array}{c} r-1 \\ \frac{1}{2}(3r-t-2) \end{array} \right)
q^{\frac{1}{2}(3r-t-2)}\, (1-q)^{\frac{1}{2}(t-r)} \nm
&=&\left( \begin{array}{c} r-1 \\ \frac{1}{2}(t-r) \end{array} \right)
q^{(r-1)-\frac{1}{2}(t-r)}\, (1-q)^{\frac{1}{2}(t-r)}\,,
\eeqa
and it is a matter of simple algebra to show that \Eq{a4.22} is a solution
of the difference equation \Eq{a4.3a}. 
This solution is valid for any spin configuration with $N_{\uparrow} = 
0, \cdots, r-1$. For the trivial cases where all spins are either up or
down, one can easily verify the obvious results 
\beqa{a4.23}
P_1^{N_{\uparrow}=r-1}(r,t=r)=1\;;\;\;\; 
P_1^{N_{\uparrow}=0}(r,t=3r-2)=1\,.              
\eeqa
An example of space-time evolution of the system based on Eq.\Eq{a4.22} is
shown in Fig.6.

\subsection{Long-time large-distance behavior}
\label{subsec:long_time}

An interesting result follows from the computation of the average time
the particle takes to cover a distance $r$ when $r$ is large.  
In \Eq{a4.22}, we set $(1-q)=p$ and $(t-r)=2a$, and we  consider
$r$ large (i.e. $r \gg 1$); then $P_1(r,t)$ can be rewritten as
\beqa{a4.24}
P_1(r,a)\,=\,\left( \begin{array}{c} r \\ a\end{array} \right) p^a\, 
q^{r-a}\,.
\eeqa
Using a standard procedure of probability theory \cite{feller1}, we have
\beqa{a4.25}
\langle a \rangle &=& \left[ \sum_a a \left( \begin{array}{c} r \\ 
a \end{array} \right) p^a\, q^{r-a} \right]_{q=1-p} 
=\left[ p\,\frac {\partial}{\partial p} (p+q)^r\right]_{q=1-p}\,=\,r\,p\,,
\eeqa
which we rewrite as 
\beqa{a4.26}
\frac{1}{2}(\langle t \rangle - r) =  r(1-q)\,,\;\;\;\mbox{or}\;\;\;
\frac{\langle t \rangle}{r} = 3 - 2q \,,
\eeqa
i.e., we retrieve the expression for the average velocity 
$c(q)=(3-2q)^{-1}$.

We now compute the explicit analytical expression of 
$P_1(r,t)$ for large~$r$. We consider $p,q$ fixed, and 
$r \gg 1$, such that $\frac {1}{r}|a-rp| \rightarrow 0$, 
which is equivalent to $[\frac {t}{r} -(3-2q)] \rightarrow 0$, the limit 
we expect for long  times. Then using Stirling's formula in the binomial 
coefficient of \Eq{a4.24}, and performing a Taylor expansion in 
$\delta a = a - rp$, \cite{feller2} we obtain 
\beqa{a4.27}
P_1(r,a)|_{r \gg 1}\,=\,(2\pi rpq)^{-1/2} \exp 
(-\delta a^2/2rpq)\,.
\eeqa
Noting that $\delta a/r = \frac {1}{2}[(t/r)-(3-2q)]$, \Eq{a4.27} 
can be rewritten as 
\beqa{a4.28}
P_1(r,t)|_{r \gg 1}\,=\,\frac {1}{\sqrt {2\pi} (rpq)^{1/2}} 
\exp -\frac {(r-\langle r(t) \rangle)^2}{2 \gamma r}\,,
\eeqa
with 
\beqa{a4.29}
\langle r(t) \rangle=t(3-2q)^{-1}=c(q)t\,,\;\;{\rm{and}}\;\; 
\gamma = 4pq (3-2q)^{-2}=4pqc^2(q).
\eeqa
In Fig.7 we show a comparison between simulation data, the computation
of the binomial expression \Eq{a4.22}, and the analytical 
result \Eq{a4.28}.

The connection with the results of subsection \ref{subsec:time_evol}, 
in particular with the function ${f}_{k=0}(t \rightarrow \infty)$, 
is easily established by taking the Fourier transform of 
\Eq{a4.28} in the limit $k=0$, that is $\int_0^\infty dr\, P_1(r,t)$;
by setting $r=x^2,  \alpha = (3-2q)/ \sqrt{8pq}\,,$ 
and $\beta = \alpha \langle r \rangle$, we obtain from \Eq{a4.28} 
\beqa{a4.29b}
\frac {1}{2} \int_0^\infty dx \,P_1(x) 
=\frac {1}{2\sqrt {2\pi}\,(pq)^{1/2}}\, e^{2 \alpha \beta} 
\int_0^\infty dx \,e^{- \alpha^2 x^2 - \beta^2/x^2} 
= \frac {1}{3-2q}\,,
\eeqa
which is exactly the result \Eq{a4.16}.\footnote{Note that the factor 
$1/2$ on the l.h.s. of \Eq{a4.29b} comes from the fact that the binomial 
coefficient in \Eq{a4.24} must be interpreted as zero if $r$ and $a$ 
are not of the same parity since $r$ and $t$ must have same parity.}

\subsection{Total probability distribution}
\label{subsec:tot_prob}

In section \ref{subsec:propag} we showed that any lattice site can be 
visited at most three times ({\em Corollary 1}). So we define $P_1(r,t), 
P_2(r,t)$, and $P_3(r,t)$, as the probabilities that the particle be at 
site $r$ at time $t$ for the first, second, and third time, respectively. 
We also introduce the probability $P_{21}(r,t)$ for the particle to be 
at site $r$ at time $t$ for the second time, without ever having visited 
the next site (at $r+1$); then the particle must have visited site $r$ 
two time-steps earlier and this was a first-time visit and the spin was 
down (see Fig.8a). Therefore
\beqa{a4.30}
P_{21}(r,t)\,=\,(1-q)P_1(r,t-2)\,.
\eeqa
Equation \Eq{a4.3a} for $f(r,t)=P_1(r,t)$ is regained by noticing that a first 
visit at site $r+1$ at time $t+1$ results from either a direct displacement 
of the particle from $r$ to $r+1$ when the spin at $r$ is up, i.e. 
$qP_1(r,t)$, or a second visit at $r$ two time-steps earlier, as described 
above; so the equation for $P_1(r,t)$ reads
\beqa{a4.31}
P_1(r+1,t+1)&=&qP_1(r,t)\,+\,P_{21}(r,t) \nm
            &=&qP_1(r,t)\,+\,(1-q)P_1(r,t-2)\,, 
\eeqa
which is exactly Eq.\Eq{a4.3a}, with $f(r,t)\equiv P_1(r,t)$.

Now to obtain an expression for $P_2(r,t)$, two cases must be considered: 
the first is that of Fig.8a, which yields the contribution given 
by \Eq{a4.30},
and the second, illustrated in Fig.8b, contributes $q(1-q)P_1(r,t-2)$. 
Consequently 
\beqa{a4.32}
P_2(r,t)\,=\,(1-q^2)P_1(r,t-2)\,.
\eeqa
There is only one situation which produces three consecutive visits to 
the same site, as shown in Fig.8c: the site must have been visited for 
the first time four time-steps earlier, and the particle must have 
moved twice backwards during these four time-steps, which yields
\beqa{a4.33}
P_3(r,t)\,=\,(1-q)^2 P_1(r,t-4)\,.
\eeqa

So the {\em total} probability that the particle be at site $r$ 
at time $t$ reads
\beqa{a4.34}
P(r,t)\,=\,P_1(r,t) +  (1-q^2)P_1(r,t-2) + (1-q)^2 P_1(r,t-4)\,,
\eeqa
with the property $\sum_r P(r,t) =1, \forall t $. 

The long-time behavior of the total probability $P(r,t)$ is readily 
obtained by taking $t \gg 1$ in \Eq{a4.34}, which yields
\beqa{a4.35}
P(r,t)|_{t \gg 1}\,=\,(3-2q)\,P_1(r,t)|_{t \gg 1}\,.
\eeqa
Since $t \gg 1$ also implies $r \gg 1$, it follows
from \Eq{a4.28} that 
\beqa{a4.36}
P(r,t)|_{t \gg 1}\,=\,\sqrt {\frac {2}{\pi}}\, \frac {1}
{(\gamma r)^{1/2}} 
\exp -\frac {(r-\langle r(t) \rangle)^2}{2 \gamma r}.  
\eeqa
where $\gamma = 4 c^2 pq, \langle r(t) \rangle = c(q) t$ (see \Eq{a4.29}).
Alternatively we write \Eq{a4.36} as
\beqa{a4.36a}
P(r,t)|_{t \gg 1}\,=\,\sqrt {\frac {2}{\pi}}\, \frac {3-2q}
{2(pq r)^{1/2}} 
\exp -\frac {(t-\la t \ra)^2}{2 (4pq) r}\,,
\eeqa
with $\la t \ra = r/c(q)$. Then the following properties, obtained 
from \Eq{a4.36} and \Eq{a4.36a},
\beqa{a4.36b}
P(r=\langle r(t) \rangle,t \rightarrow \infty) 
= \frac{1}{\sqrt{2 \pi}}\left( \frac{(3-2q)^3}{q(1-q)}\right)^{1/2}
\frac{1}{t^{1/2}}\,,  
\eeqa
and 
\beqa{a4.37}
\la (t - \la t \ra)^2 \ra^{1/2}\,=\,2 \sqrt {2pq}\, r^{1/2}
\eeqa
show that , for long times, the amplitude of the probability distribution 
$P(r,t)$ decays like $t^{-1/2}$ and its width grows like $\sqrt {r}$. 

We also note that, by taking the sum over $r$ in \Eq{a4.34}, followed by 
the long-time limit ($t \gg 1$), we obtain
\beqa{a4.38}
1\,=\, (3-2q) \lim_{t \rightarrow \infty} \sum_r P_1(r,t)\,,\;\;\;
\mbox{or} \;\;\; \lim_{t \rightarrow \infty} \sum_r P_1(r,t)\,=\,c(q)\,,
\eeqa
in accordance with our previous results (see subsection 
\ref{subsec:time_evol}). In Fig.7 we show the long-time behavior of
$P_1(r,t)$, which is the same (within a constant factor, see \Eq{a4.35})
as that of $P(r,t)$.

Finally it is interesting to note that $P(r,t)$, as given by \Eq{a4.36a}, 
is the solution of the differential equation
\beqa{a4.39}
\partial_r P(r,t) \,+\, \frac{1}{c}\,\partial_t P(r,t) &=&\frac{1}{2}\, 
(4pq)\, \partial_t^2 P(r,t)\,, 
\eeqa 
which can be shown to be the continuous limit of the difference equation 
for $P(r,t)$ \cite{boon}.  The appearance of a Gaussian distribution 
in time, \Eq{a4.36a}, and the differential equation with propagation 
\Eq{a4.39} for long times are natural consequences of the averaging 
over the initial spin distributions.

\newsec{The 2-D triangular lattice}
\label{sect:tri_lattice}

We consider the motion of a particle on a triangular lattice fully 
occupied by  {\em rotators} (flipping scatterers)  which rotate the 
velocity vector of the particle by an angle of $2\pi/3$ either to the 
right or to the left, depending on whether the rotator on the site hit by 
the particle is a R or a L scatterer, respectively.\footnote{Although 
the spin notation is applicable to the rotators (with R $\equiv \uparrow$ 
and L $\equiv \downarrow$), the right (R) and left (L) characterization 
of the scatterers, as used in \cite{cohenetal}, is more convenient for 
our discussion of the triangular lattice.} The state of the rotator at the 
site where the scattering took place, changes: R~$\Leftrightarrow$~L.
Simulations show that, whatever the initial configuration of scatterers,
the particle always goes into a propagation phase in the direction of one
of the lattice axes \cite{cohenetal}. An example is given in Fig.9. Here
we show how propagation can be proved mathematically, and we derive the
statistical properties of the dynamics of the particle.

\subsection{Definitions}
\label{subsec:definitions}

\noindent (i) The propagation of the particle on the triangular lattice 
always takes place in a {\em strip} which is defined a follows: 
a strip is a region of the triangular lattice bounded by two adjacent 
(parallel) lines, both oriented along one of the lattice axes.

\noindent (ii) In accordance with the 1-D case, we define particle 
propagation as follows: a particle propagates in one direction in
a strip if its motion is confined to the  strip (defined by its initial
velocity and its velocity after the first scattering event) where the
particle visits any site not more than a fixed number of times.

\noindent (iii) The particle's velocity vector at time $t$ is denoted 
by ${\bf C}(t)$.

\subsection{The propagation theorem}
\label{subsec:prop_theor}

\noindent {\em Theorem 2} : For any initial distribution of scatterers 
on a triangular lattice, a moving particle propagates in one particular 
direction through a strip on the lattice; this strip and the direction of 
propagation along it depend on the initial configuration of the scatterers
at the origin (the initial position of the particle) and at three
neighboring sites; these four sites form a parallelogram whose orientation
determines the propagation strip. \\

\noindent {\em Proof} : 
First we show that for each possible  initial scatterers configuration, a 
propagation of the particle will occur in one of four strips, as sketched in 
Fig.10 (propagation directions and strips are indicated by F(forwards), 
U(upwards), D$_1$ and D$_2$(downwards)). The strip in which the particle will 
propagate (F, U, D$_1$ or D$_2$) depends on the initial velocity direction 
${\bf C}(t_0)$, and on the initial state (R or L) of the four sites forming 
a parallelogram which contains ${\bf C}(t_0)$. There are four such 
parallelograms, hence four possible directions (see Fig.10): 

\1\ \2\ \3\ \6\ $\rightarrow$ F, \ \1\ \2\ \4\ \7\ $\rightarrow$ U, \ 
\1\ \2\ \3\ \8\ $\rightarrow$ D$_1$, \ \1\ \2\ \4\ \5\ $\rightarrow$ D$_2$.\\
Then we demonstrate that once the particle is inside a strip (at one of 
the sites indicated by \ \first\ in Fig.10), it never leaves this strip.
Like in 1-D, the key ingredient of the ultimate unidirectional propagation
of a particle on a triangular lattice fully occupied with flipping (R 
and L) scatterers, is the existence of a blocking mechanism.
On the triangular lattice, this mechanism produces a {\em blocking pattern}
based on a {\em zig-zag} path of length four (in lattice unit lengths), 
which contains parallel velocity vectors at even or at odd times (see 
examples in Fig.11), and which prohibits particle motion 
more than one lattice unit length in the direction opposite to 
the propagation direction; as a consequence any site visited by
the particle cannot be visited more than six times (see
{\em Corollary 2} below). The result is that the particle propagates in 
a strip in the direction naturally defined by the blocking pattern. 

Assume that a blocking pattern has been formed at the sites visited by 
the particle at times $t, t+1, t+2$ and $t+3$, so that the vectors 
${\bf C}(t+)$ and ${\bf C}((t+2)+)$, and ${\bf C}((t+1)+)$ and 
${\bf C}((t+3)+)$ are parallel, respectively (see Fig.11a).
Then such a zig-zag trajectory can continue in two ways:
either the particle  continues directly its zig-zag motion at the next time 
step (see Fig.12a), or it turns back (Fig.12b).  Obviously, the direct
continuation of the zig-zag trajectory leads to propagative motion. So in 
order to prove propagation, we only have to consider the result of the turn 
back motion, and show that the particle nevertheless continues to move 
along the strip defined by the zig-zag direction along which it was moving 
from time $t$ to time $t+3$. Now when the particle turns back, its 
trajectory creates a triangle (Fig.12b) where from a zig-zag path emerges 
(Fig.12c). Propagation then occurs as a consequence of the 
fact that backward motion at time $t+4$, necessarily produces a blocking 
pattern, which forbids the particle to move more than one lattice unit length 
in the backward direction along the edge of the strip; as a result the 
particle will ultimately move one step further along the same strip in the 
zig-zag direction in which it was moving forward at time $t+3$.
Therefore any trajectory eventually forms a zig-zag of length 4, 
leading to propagation in the direction of the zig-zag path.
One can verify that after the turning back of the particle,
the trajectory forms a parallelogram (Fig.12c). Now the path of 
the particle through this parallelogram ends with three steps with 
velocity vectors: ${\bf C}((t+7)+), {\bf C}((t+8)+), {\bf C}((t+9)+)$ 
that together form a zig-zag pattern of length 3 (see caption of Fig. 12).  
But these three velocity vectors form together with ${\bf C}((t+10)+)$ a 
new zig-zag of length 4, i.e. a new blocking pattern (such as in Fig.12a), 
shifted one lattice unit length in the direction of propagation, with 
respect to the initial blocking pattern. The cases shown in the figures 
used for the demonstration are typical and all other cases are  easily  
inferred from these typical situations, as the reader can verify.
Consequently, we have shown that a blocking pattern always gets shifted 
by one lattice unit length along an edge of the strip in the direction of 
propagation, either after two time steps if the previous zig-zag motion is 
continued (Fig.11a), or after seven time steps if the trajectory is 
turned back and a parallelogram of motion is formed (Fig.12c).
  
Thus, the particle will always propagate along a strip in the 
direction determined by the blocking mechanism in this strip. 
For an (arbitrary) given initial velocity of the particle, propagation 
can only take place in one of four strips: two with angles $\pm \pi/3$ 
(strips F and U) and two with angles $\pm \pi$ (strips D$_1$ and D$_2$) 
with respect to the initial velocity direction.  
The ratios of the occurrence of propagation along the four strips can be 
evaluated from Fig.10 by examining all possible paths leading to a strip (see 
examples in Fig.11) and by giving weights $q$ and $(1-q)$, to an R and an L 
scatterer, respectively. For $q= \frac{1}{2}$, these ratios are 3:3:1:1
for the directions F, U, D$_1$ and D$_2$ respectively.

\bigskip

Corollaries follow from Theorem 2.\\

\noindent {\em Corollary 1} : The shortest time to create a blocking 
pattern is four time steps, and the longest time is ten time steps.\\
{\em Proof} : The proof follows straightforwardly by inspection of Figs.12a 
and 12c, where the zig-zag pattern of four velocities arranged in a W shape 
forms after 4 and 10 successive displacements of the particle, respectively
(as verified by counting the number of arrows after the initial state).\\

\noindent {\em Corollary 2} : The maximum number of visits to any site 
visited by the particle on the triangular lattice is six, and the 
maximum number of passages along any link connecting two neighboring 
sites is five. \\
{\em Proof} :  Fig.12c. shows the longest trajectory executed by the 
particle to move one unit length along the lattice  in the 
propagation strip. When the scatterer configuration is such that the 
particle is again forced into a turn back motion (as in Fig.12b), when 
visiting the next site on the strip, the parallelogram of Fig.12c 
repeats itself (shifted upside down by one lattice unit length in the 
propagation direction). Then the middle site of the parallelogram, which
has been visited three times (site \ \6\ in Fig.12c) is visited three more 
times, yielding a total number of six visits. The number of visits to a site 
is then easily evaluated by counting the number of incoming arrows pointing 
to the site. For instance, in Fig.12c, there are three incoming arrows to 
site \ \6\ , and, when at the next time step, the particle undergoes 
backward motion thereby producing again the parallelogram trajectory (shifted 
upside down to the right), the three arrows pointing towards site \ \3\ are 
now pointing to site \ \6\ , which yields a total number of six visits to 
site \ \6\ . This is the maximum number of visits since the blocking 
mechanism prevents the particle from ever coming back to this site. 
Similarly the number of passages along a link between two neighboring sites 
is evaluated by counting the number of arrows on that link. In the above 
example, which shows the typical case where the particle undergoes the 
longest possible trajectory to proceed in the propagation strip, the number 
of passages along link \ \3\-- \6\ is $3+2=5$. \\

\noindent {\em Corollary 3} : All visited sites located on one edge of the 
propagation strip are in the same state (R or L), and all sites located on 
the other edge are in the opposite state (L or R, respectively). \\
{\em Proof} : The blocking mechanism produces a zig-zag pattern with 
alternating velocity vectors (see Figs.11 and 12); as a result the 
states of the visited sites must be alternately R and L, or L and R.
Consequently, the propagation dynamics triggers a reorganization of the 
states of the scatterers of the visited sites.

\subsection{Statistical properties} 
\label{subsec:2D_stat_prop}

We characterize the propagation process by the increase in the value of
the coordinate of the particle along one edge of the propagation strip,
as illustrated in Fig.13. We consider the situation where the particle
visits site $r$ for the first time, and we analyze how the
particle proceeds from site $r$ to the (next) site with coordinate $r+1$ 
on the same side of the strip. Obviously, before arriving at site $r+1$,
the particle must go through a site with coordinate $r' = r+ \frac{1}{2}$ 
on the other side of the strip (see Fig.13).  At both site $r$ and site 
$r+\frac{1}{2}$, one of two possiblities occurs: either the particle 
trajectory continues a forward zig-zag path or it turns back.  Thus, 
the transition from $r$ to $r+1$ can take place in four possibile ways: \\
($ff$) the particle continues a forward zig-zag path at both sites $r$ and 
$r+\frac{1}{2}$; \\
($fb$) the particle continues a zig-zag path at $r$ and  turns back at
$r+\frac{1}{2}$; \\
($bf$) the particle turns back at $r$, and goes into a forward zig-zag path 
at $r+\frac{1}{2}$; \\
($bb$) the particle turns back both at $r$ and at $r+\frac{1}{2}$.

So, in case ($ff$) the transition from $r$ to $r+1$ occurs in two forward 
time steps (see Fig.12a), and obviously takes longer in the other cases. 
Indeed, in section \ref{subsec:prop_theor}, we showed that if the particle 
turns back (see Fig.12b), it takes seven (6+1) time steps before it performs 
a zig-zag path again along the propagation strip (see Fig.12c).  Therefore, 
the transition takes eight times steps in case ($fb$) -- (1+7) corresponding 
to one step forward and one back turn -- and (7+1) time steps in case ($bf$), 
while  fourteen time steps are necessary in case ($bb$) -- (7+7) 
corresponding to two back turns. It is easy to verify that the occurence 
probabilities of the various cases are: $q(1-q)$ for case ($ff$), $q^2$
for case ($fb$), $(1-q)^2$ for case ($bf$), and $(1-q)q$ for case ($bb$).

From the above results, we can write the equation for the first visit
probability at site $r+1$ at time $t$  as
\beqa{a5.4}
P_1(r+1,t)&=& [q^2 +(1-q)^2] P_1(r,t-8)  \nm
          &+& q(1-q) [P_1(r,t-2) + P_1(r,t-14)]\,.
\eeqa
It follows from this expression that the average time it takes the particle 
to propagate one lattice unit length along a side of a strip is
$\langle t \rangle = 8\, [q^2 +(1-q)^2] + 16\, q(1-q) = 8\, 
\mbox{time steps}$, so that the average propagation velocity reads
\beq{a5.5}
\langle c \rangle = \frac{1}{\langle t \rangle} = \frac{1}{8}\,,
\eeq
(in lattice unit lengths per time step), independently of $q$
and of the direction of propagation.

When the particle is propagating in a strip, it takes an even number of 
time-steps to move from one site to the next on the edge of the strip 
in the forward direction (e.g. from $r$ to $r+1$ via $r'$ in Fig.13) 
because of the blocking zig-zag pattern.\footnote{A backward displacement
(in the direction opposite to the propagation direction) can only take
place along the edge of the strip and takes one time step since, because
of the blocking mechanism, it cannot exceed one lattice unit length.}
Therefore when the particle arrives {\em for the first time} at site $r+1$ 
(see Fig.13), site~$r$ cannot have been visited more than four times (as 
can be checked with Fig.12); so we can also write Eq.\Eq{a5.4} as

\beq{a5.5a}
P_1(r+1,t) = 
q(1-q) P_1(r,t-2) + \sum _{\alpha = 2}^4 P_{\alpha}(r,t-2)\,,
\eeq
where, as can be verified using Fig.13 to explore all possible paths,
\beqa{a5.5b}
P_2(r,t-2)&=&(1-q)^2 P_1(r,t-8)\,, \nm
P_3(r,t-2)&=& q^2 P_1(r,t-8)\,, \nm
P_4(r,t-2)&=&(1-q)q \,P_1(r,t-14)\,.
\eeqa
Now we define the total probability 
$P(r,t)=\sum _{\alpha = 1}^4 P_{\alpha}(r,t)$, which, with (\ref{a5.5b}),
is given by
\beqa{a5.5c}
P(r,t) \,=\, P_1(r,t) + [q^2 + (1-q)^2] P_1(r,t-6) + q(1-q) P_1(r,t-12)\,.
\eeqa
In \Eq{a5.5c}, we use  equation \Eq{a5.4} for $P_1(r,t')$ with 
$t'=t, t-6, t-12$, and in the result, we combine the various terms to
obtain a closed equation for the total probability 
\beqa{a5.5d}
P(r,t) \,=\, q(1-q)[P(r-1,t-2) + P(r-1,t-14)] \\ 
+ [q^2 + (1-q)^2]P(r-1,t-8)\,.
\eeqa
We observe that $P(r,t)$ obeys the same equation as $P_1(r,t)$, 
Eq.\Eq{a5.4}, and it can be shown \cite{boon} that its 
continuous limit has the same structure as Eq.\Eq{a4.39}.

\subsection{Long-time behavior}
\label{subsec:2D_long_time}

To investigate the long-time behavior of the particle motion, we proceed
along the same lines as in the one-dimensional case. We start with the
Laplace-Fourier transformation of Eq.\Eq{a5.4}, which yields
\beqa{a5.6}
e^{{\imath}k} \left[e^{14 s} \tilde{f}_k(s) - 
\sum_{m=1}^7 e^{2m}s f_k(t=14-2m) \right]&=&   \nm
\left[ q^2 + (1-q)^2 \right] \left[ e^{6s} \tilde{f}_k(s) 
- \sum_{m=1}^3 e^{2m}s f_k(t=6-2m) \right]&+& \nm
q(1 - q) \left[ (1 + e^{12 s}) \tilde{f}_k(s) 
- \sum_{m=1}^6 e^{2m}s f_k(t=12-2m) \right]&,&  
\eeqa 
where $f_k(t)$ denotes the space-Fourier transform of $P_1(r,t)$, and
$\tilde{f}_k(s)$ its Laplace transform. In the limit $k=0$, we obtain 
\beqa{a5.7}
\tilde{f}_{k=0}(s)\,=\,\frac{h(s)}{g(s)} 
\eeqa
with 
\beqa{a5.8}
h(s)=\left[\sum_{m=1}^7 e^{2ms}{f}_{k=0}(14-2m)\right]  
- q(1-q)\left[\sum_{m=1}^6 e^{2ms}{f}_{k=0}(12-2m)\right]  \nm
-[q^2 + (1-q)^2]\left[\sum_{m=1}^3 e^{2ms}{f}_{k=0}(6-2m)\right],
\eeqa
and
\beqa{a5.9}
g(s)=e^{14s}-q(1-q) e^{12s}-[q^2+(1-q)^2] e^{6s}- q(1-q).
\eeqa
To solve Eq.\Eq{a5.7}, we need to know the ``initial conditions'', i.e.
the values of $f_k(t)$ at $t=2m$, for $m=0, \cdots, 6$. The values of
$f(r,t)$ at $t=0,\cdots, 12$ are obtained from the particle dynamics 
discussed in the previous section and are given in Appendix B, along 
with their corresponding Fourier transforms.  We insert these values
in \Eq{a5.8}, and, since we are interested in the long-time behavior,
we consider the limit $\tilde{f}_{k=0}({s \rightarrow 0})$
by expanding $\tilde{f}_{k=0}(s)$ to lowest significant order in $s$;
the result is
\beqa{a5.10}
\lim _{s \rightarrow 0} \tilde{f}_{k=0}(s) = \frac{1}{8s}\,,
\eeqa
which, by inverse transformation, yields
\beqa{a5.11}
\lim _{t \rightarrow \infty} f_{k=0}(t) = \frac{1}{8} = 
\langle c \rangle \,.
\eeqa

Thus, as for the one-dimensional case, we find that in the long-time
limit, $f_{k=0}(t\rightarrow \infty) \equiv \lim _{t \rightarrow \infty}
\sum_r P_1(r,t)$ is equal to the average propagation velocity, with
the difference that, in the triangular lattice, $\langle c \rangle$
is independent of~$q$. Furthermore it can be shown \cite{boon} that the 
long-time solution  of Eq.\Eq{a5.4} is given by 
\beqa{a5.12}
P_1(r,t)|_{t \gg 1}\,=\,\sqrt {\frac {2}{\pi}}\, \frac {1}
{({\tilde \gamma}r)^{1/2}} 
\exp -\frac {(t-\la t \ra)^2}{2 {\tilde \gamma} r}\,,
\eeqa
with $\la t \ra = r/{\la c \ra} = 8r$ and ${\tilde \gamma} = 72 pq$.
In Fig.14 we present numerical simulations and show that our 
theoretical result \Eq{a5.12} is in agreement with the simulation data.

\section{Concluding comments}
\label{sec:concl_comm}

\noindent 1. By replacing spins $\uparrow$ and $\downarrow$ by 1's 
and 0's respectively, the initial state of the 1-D spin lattice can be 
read as the input string of the tape of a Turing machine whose control 
device is the particle; the controller performs operations according to 
the rules of the flipping spins and writes the output on the tape as the 
final spin configuration also converted into 1's and 0's.  Here the 
computation algorithm performs the binary addition of the initial string 
with itself followed by a ``XOR'' with 1 (which is equivalent to 1 
$\Leftrightarrow$ 0 plus a shift). In the triangular lattice, one can
similarly interpret the state of the scatterers along the edges of the
propagation strip as R/L $\rightarrow$ 0/1; then the algorithm performs
the logical operations ``AND'' and ``NAND'' with 0 alternately on each
side of the strip, which transforms a random sequence of 0's and 1's 
into a periodic string ...01010101... .  \\

2. After the  passage of the particle through the one-dimensional 
lattice, all the spins are in the state opposite to their initial state
($\uparrow \Leftrightarrow \downarrow $) with one lattice site shift in
the direction opposite to the propagation direction. Consider that the
the system is made periodic (i.e. on a circle) by identifying the first 
and last sites of the chain, and that the particle propagates clockwise
on the circle; then there is a counterclockwise drift (two lattice sites 
back) of the initial spin configuration every two cycles, and the average
propagation velocity is given by $c(q)=\frac{1}{2}[\frac{1}{3-2q}+
\frac{1}{3-2(1-q)}]$. Similarly, when, in the triangular lattice, one 
attaches the two beginning and two end sites of a strip crosswise,
the particle undergoes periodic motion on the strip. \\

3. Propagation has also been observed in the square lattice fully
occupied with flipping scatterers when the scatterers are distributed 
periodically over the lattice \cite{cohenetal}. However, not in all such 
cases does propagation occur during the time span of the simulations
($\leq 10^7$ time steps). It remains therefore an open question whether
propagation on the square lattice, periodically covered with flipping
scatterers, always occurs or only for a certain class of periodic
scatterer patterns. When it occurs, propagation seems always to
take place in a strip (like on the triangular lattice), which shows
similarity with the glider behavior in the Game of Life \cite{game}. 
Although the precise blocking mechanism is not known for the square 
lattice, ``immediate'' propagation can occur, which would imply then the
possibility of an ``instantaneous'' blocking on the square lattice. \\

4. Propagation on the one-dimensional and triangular lattices occurs
for scattering angles of $\pm \pi$ and $\pm 2\pi/3$, respectively.
For those cases on the periodically occupied square lattice
where propagation has been found, the scattering angle equals $\pm \pi/2$, 
while for the randomly occupied square lattice, the maximum angle 
scattering rule yields a(n) (average) value larger than $\pi/2$.  On the 
other hand, a scattering angle of $\pm \pi/3$ on the triangular lattice 
leads to the motion of the particle on a honeycomb (hexagonal) lattice 
where no propagation seems to occur for flipping scatterers \cite{cohenetal}.
One might therefore conjecture that propagation only occurs in lattices 
fully occupied with flipping scatterers, when the scattering angle is 
$\geq  \pi/2$ in absolute value.\\

5. Notice that in the development performed for the triangular
lattice, we have not used the regularity property of the Bravais lattice.
So the results should not be dependent of this property and should also 
be valid for the random triangular lattice such as the Delaunay lattice 
(see Fig.1), provided the scattering rule is set such that the particle 
is deflected with the largest possible angle ($> \pi/2$). However, because
of the topological randomness of the Delaunay lattice, one cannot predict 
whether the trajectory of the particle will remain unbounded. \\

6. To the best of our knowledge, it is an open question whether
particle dynamics in three dimensional lattices with flipping scatterers
can exhibit propagation or other peculiar behaviors. \\

\newpage

\renewcommand{\theequation}{\Alph{chanum}.\arabic{eqnnum1}}
\setcounter{chanum}{1}
\setcounter{eqnnum1}{0}

\noindent{\large{\bf{Appendix A}}}\\

Here we derive the expression for the propagation velocity of a particle 
moving in a 1-D lattice with flipping spins, directly from the particle 
dynamics described in section \ref{subsec:basic_eqs}. The average
asymptotic propagation velocity is given by 
\beqa{A.1}
\la c \ra \,=\,\lim_{t \gg \Delta t} C(t)\,,
\eeqa
with
\beqa{A.2}
C(t)\,=\,\frac {R(t) - R(0)}{t}\,,
\eeqa
and where $\Delta t$ is the unit time step ($\Delta t = 1$, i.e. the limit
in (A.1) will be taken as $t \rightarrow \infty$). 
It follows by iteration from Eq.\Eq{a2.1}, that
\beqa{A.3}
R(t) - R(0)\,=\,C(t-1) + C(t-2)) + \cdots +C(0)\,=\,
\sum_{\tau=0}^{t-1}C(\tau)\,.
\eeqa

When we consider the velocity of the particle at two successive
time-steps,  five cases must be examined as described in Table 1
(see also Fig.15). If the particle arrives at a site at time $\tau-1$ 
with velocity $C(\tau -1)$ and leaves
the site with velocity $C(\tau )$, we observe that when taking the
sum over the velocities, all $C(\tau -1)$'s and $C(\tau )$'s which do not 
cancel are those which correspond to particle displacements leading 
to a site visited by the particle for the first time. 
So the sum $\sum_{\tau}C(\tau)$ is equal to 
the number $N_f$ of displacements which, during time $t$, lead to a 
{\em first visit}. Now, since the particle undergoes a displacement at
each time-step, the total number of displacements $N_t$ during 
time $t$, is equal to $t$. Consequently, from (A.1)-(A.3), we have
\beqa{A.4}
\la c \ra \,=\,\lim_{t\rightarrow \infty }\frac {1}{t}\,[R(t) - R(0)]\,=\,
\lim_{t\rightarrow \infty }\frac {1}{t}\,\sum_{\tau=0}^{t-1}C(\tau) 
\,=\,\lim_{t\rightarrow \infty }\frac {N_f(t)}{N_t(t)}\,;
\eeqa
and, since $N_f(t)/N_t(t)$ is just $\sum_r P_1(r,t)$, it follows that 
\beqa{A.5}
\la c \ra \,=\,\lim_{t\rightarrow \infty }\sum_r P_1(r,t)
\,=\,\lim_{t \rightarrow \infty} {f}_{k=0}(t)
\,=\,\frac{1}{3-2q}\,\equiv c(q), 
\eeqa
where we have used (3.16).

\bigskip

\begin{tabular}{|l|l|l|l|l|}  \hline
\multicolumn{5}{|c|}{TABLE 1} \\ \hline\hline
{ } &
{The particle } &
{with} &
{and leaves } &
{The next site} \\
{Fig.} &
{arrives at site} &
{velocity} &
{site $r$ with} &
{is visited} \\ 
{15} &
{$r$ for the} &
{$C(\tau -1)$}   &
{velocity $C(\tau)$} &
{for the} \\ \hline\hline
(a) & first time  & $C=+1$ &  $C=+1$  & first time  \\ 
(b) & first time  & $C=+1$ &  $C=-1$  & second time \\
(c) & second time & $C=+1$ &  $C=+1$  & first time  \\ 
(d) & second time & $C=-1$ &  $C=+1$  & second time \\
(e) & third  time & $C=-1$ &  $C=+1$  & second time \\
\hline
\end{tabular}

\vspace{.25in}

\noindent{\large{\bf{Appendix B}}}

\vspace{.25in}

\setcounter{chanum}{2}
\setcounter{eqnnum1}{0}

The ``initial conditions'' for the computation of the long-time behavior
of the function $f_{k=0}(t) \equiv \sum_r P_1(r,t)$ in section 4.4,
are given by the following expressions 
\beqa{B.1}
f(r,t=0)&=&\delta_{r,0}\,, \nm
f(r,t=2)&=&q(1-q)\,\delta_{r,1}\,, \nm
f(r,t=4)&=&q^2(1-q)^2\,\delta_{r,2}\,, \nm
f(r,t=6)&=&q^3(1-q)^3\,\delta_{r,3}\,, \nm
f(r,t=8)&=&q^4(1-q)^4\,\delta_{r,4}\,+\,[q^2+(1-q)^2]\,\delta_{r,1}\,,\nm 
f(r,t=10)&=&q^5(1-q)^5\,\delta_{r,5}
\,+\,2q(1-q)[q^2+(1-q)^2]\,\delta_{r,2}\,,\nm 
f(r,t=12)&=&q^6(1-q)^6\,\delta_{r,6}
\,+\,3q^2(1-q)^2[q^2+(1-q)^2]\,\delta_{r,3}\,,
\eeqa 
and correspondingly, in Fourier transform, by
\beqa{B.2}
f_k(0)&=&1\,, \nm
f_k(2)&=&q(1-q)\,e^{\imath k}\,, \nm
f_k(4)&=&q^2(1-q)^2\,e^{2\imath k}\,, \nm
f_k(6)&=&q^3(1-q)^3\,e^{3\imath k}\,, \nm
f_k(8)&=&q^4(1-q)^4\,e^{4\imath k}\,+\,[q^2+(1-q)]^2\,e^{\imath k}\,, \nm
f_k(10)&=&q^5(1-q)^5\,e^{5\imath k}
\,+\,2q(1-q)[q^2+(1-q)^2]\,e^{2\imath k}\,, \nm
f_k(12)&=&q^6(1-q)^6\,e^{6\imath k}
\,+\,3q^2(1-q)^2[q^2+(1-q)^2]\,e^{3\imath k}\,.
\eeqa

\newpage

{\Large{\bf{Acknowledgments}}}

\bigskip

PG benefited from a stay at Rockefeller University whose hospitality is
gratefully acknowledged.
JPB acknowledges support by the {\em Fonds National de la Recherche 
Scientifique} (FNRS, Belgium). EGDC gratefully ackowledges support
from the {\em Belgian Franqui Foundation} and the hospitality of the 
{\em Center for Nonlinear Phenomena and Complex Systems} at the 
Universit\'{e} Libre de Bruxelles, which made an extended stay in 
Brussels possible, during which a collaboration with PG and JPB initiated 
this work. EGDC also acknowledges support from the US Department of Energy 
under grant number DE-FG02-88-ER13847. LAB was partially supported by
NSF grant number DMS-963063. EGDC and LAB thank the {\em Erwin 
Schr{\"o}dinger Institute} (Vienna) for its hospitality and support.

\newpage

\noindent{\Large \bf Figure captions}

\bigskip

\begin{list}{}{}
\item 
{\underline{Fig. 1.}}  Propagation in a random Delaunay lattice, 
where the particle arriving at a site of the random lattice is
deflected over the largest possible angle there, either to the right 
or to the left, depending on the R or L nature of the scatterer.
Arrows illustrate particle displacements. The shaded area shows the
propagation strip.\\
\item 
{\underline{Fig. 2.}} Scattering on the one-dimensional lattice 
illustrating the proof of {\em Theorem 1} (see text); the blocking patterns 
are framed with dotted squares.\\
\item
{\underline{Fig. 3.}}  Example of spin reorganization in 1-D lattice. 
The broken line is the particle trajectory. Black dots are sites with spin 
down, and open squares (shown only for the initial and final configurations) 
are sites with spin up.  The spin reorganization is well observed by
comparing the top row with the bottom row: 
$\bullet \Leftrightarrow \Box$ (with one site shift to the left). \\
\item
{\underline{Fig. 4.}}  Propagation velocity in 1-D lattice: for $q=1$ (all
spins initially up) $\la c \ra = 1$ (upper line): $q=0.5$ (for two different 
spin configurations; middle lines) $\la c \ra = 1/2$; and $q=0$ (all spins 
initially down) $\la c \ra = 1/3$ (lower line). \\
\item
{\underline{Fig. 5.}}  $f_{k=0}(t)$ as a function of $t$ for $q=0.1$.
Simulation results (dots) and theory, Eq.\Eq{a4.12} (lines).\\
\item
{\underline{Fig. 6.}}  Space-time evolution of $P_1(r,t)$ for $q=0.5$.
The x-axis denotes space.\\
\item
{\underline{Fig. 7.}}  The probability $P_1(r,t)$ as a function of
$r$ at $t=500$. Simulation data (black dots), binomial expression \Eq{a4.22}
(open circles), and Eq.\Eq{a4.28} (curve).\\
\item
{\underline{Fig. 8.}}  Situations on the 1-D lattice leading to second visits,
(a) and (b), and to third visit, (c).\\
\item
{\underline{Fig. 9.}} Typical example of  propagation pattern in 
triangular lattice. The initial position of the particle is shown by 
a full diamond; the propagation direction is along the upper left axis
of the lattice. The coordinates are marked in lattice unit lengths. \\ 
\item
{\underline{Fig. 10.}}  Propagation strips on the triangular lattice.
The initial velocity of the particle, ${\bf C}(t_0)$, is shown as the 
heavy arrow. When the particle arrives at one of the sites marked \ \first\ , 
it is trapped in one of the propagation strips bounded by parallel heavy 
solid lines. The four propagation directions are indicated by arrows and 
capital letters (F, U, D$_1$, D$_2$). \\
\item
{\underline{Fig. 11.}}  Examples of trajectories leading to propagation
in one of the four directions (F, U, D$_1$ or D$_2$). 
Light arrows indicate the successive velocity vectors of the particle 
after the initial state (${\bf C}(t_0)$, heavy arrow).\\
\item
{\underline{Fig. 12.}} Formation of blocking pattern by direct forward
zig-zag path (a), and by turn back motion (b) followed by zig-zag
trajectory emerging from parallelogram path (c). Light arrows indicate 
the successive velocity vectors of the particle. Vectors ${\bf C}((t+7)+), 
{\bf C}((t+8)+), {\bf C}((t+9)+)$, and  ${\bf C}((t+10)+)$ are along the 
links \ \2\-- \3\ , \ \3\-- \6\ , \ \6\-- \first\ , and \ \first\--$\,$ \0\ , 
respectively. \\
\item
{\underline{Fig. 13.}} Particle motion from site $r$ to site $r+1$
along the propagation strip {\em via} site $r'= r + \frac{1}{2}$. The
arrow pointing to the right shows the propagation direction. \\
\item
{\underline{Fig. 14.}} Long-time behavior in triangular lattice.
$P_1(r,t)$ as a function of $r$ after 7616 time-steps: numerical data 
from simulation measurements (black dots) compared with analytical result 
\Eq{a5.12} (full curve). \\
\item
{\underline{Fig. 15.}} Illustration of the five cases discussed in
Appendix A and described in Table 1.

\end{list}




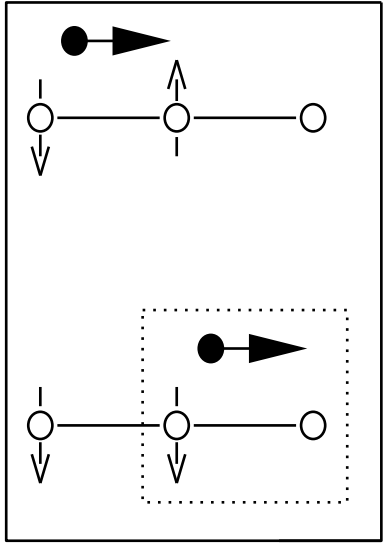
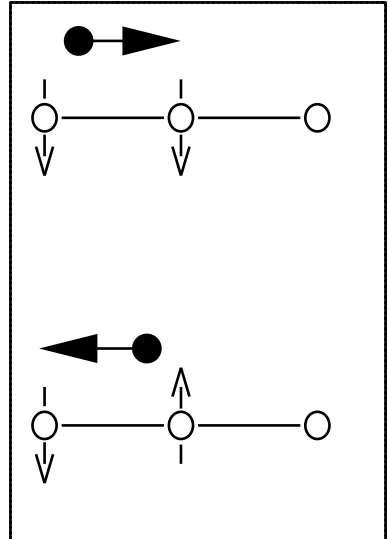
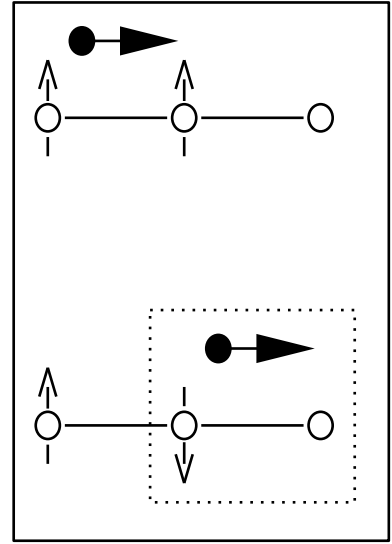
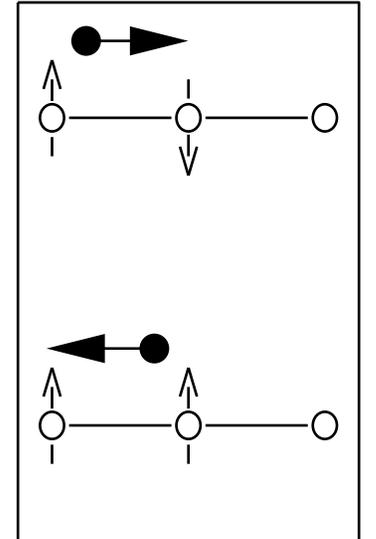
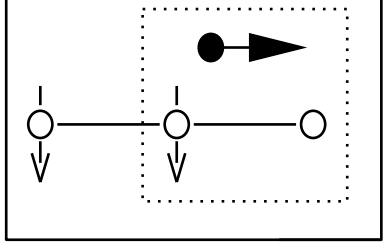
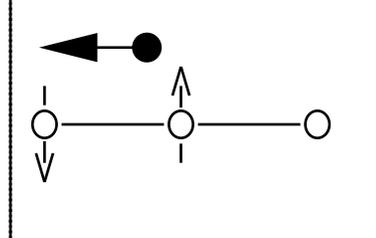
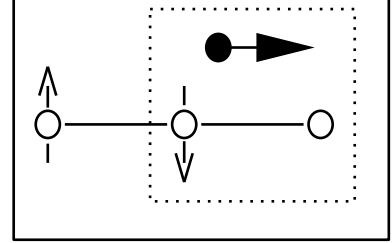
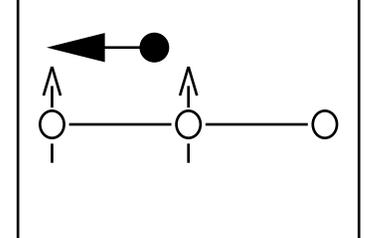
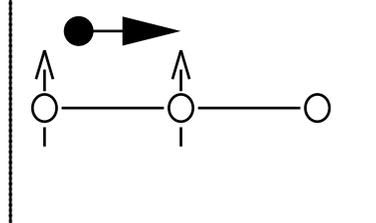
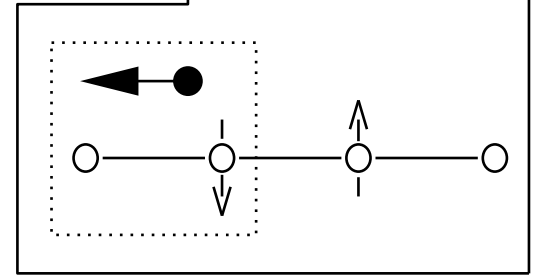
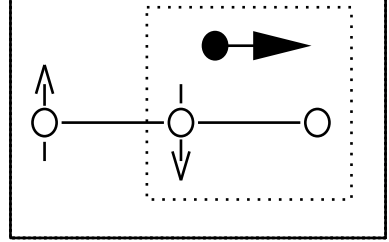
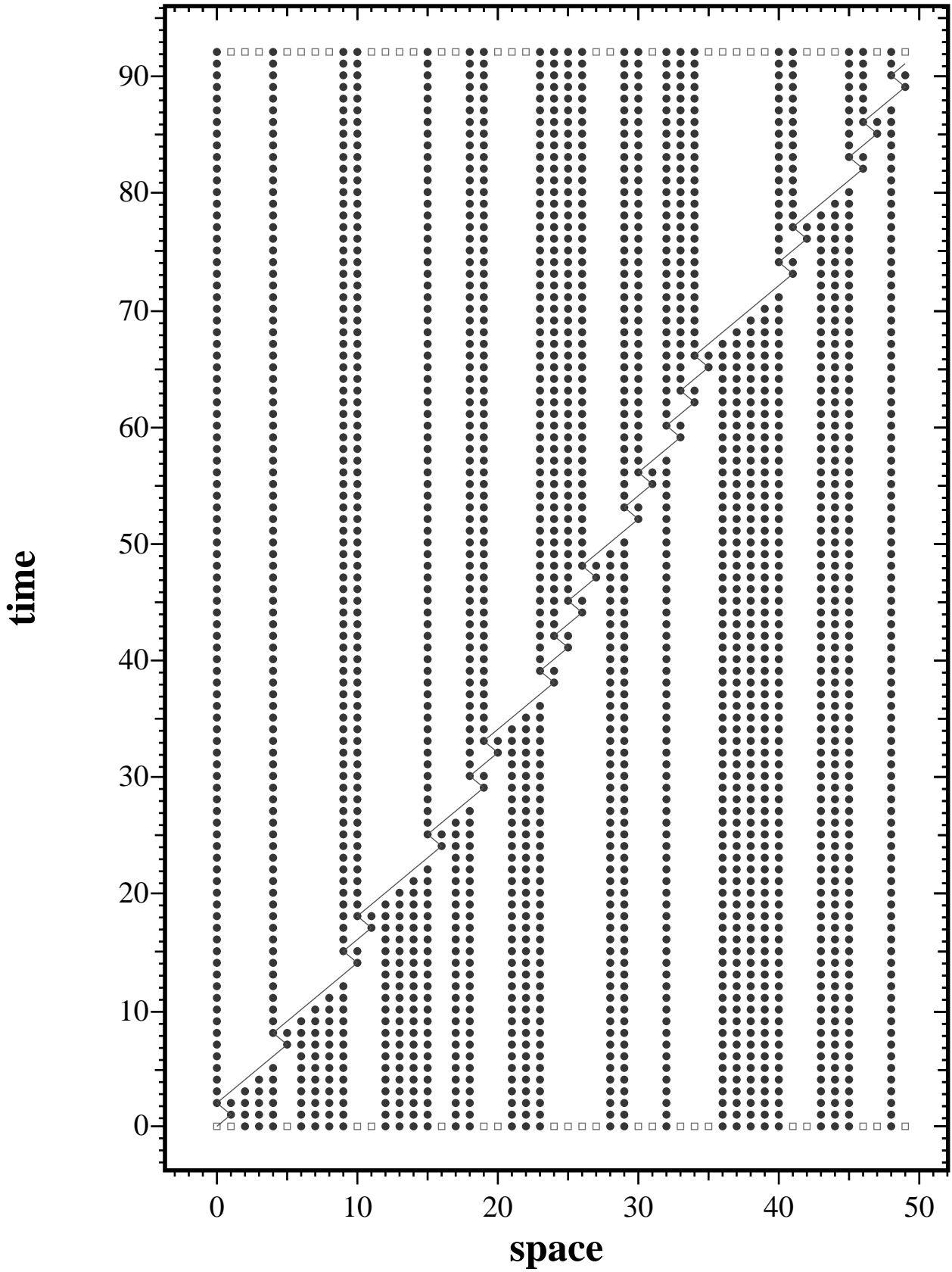
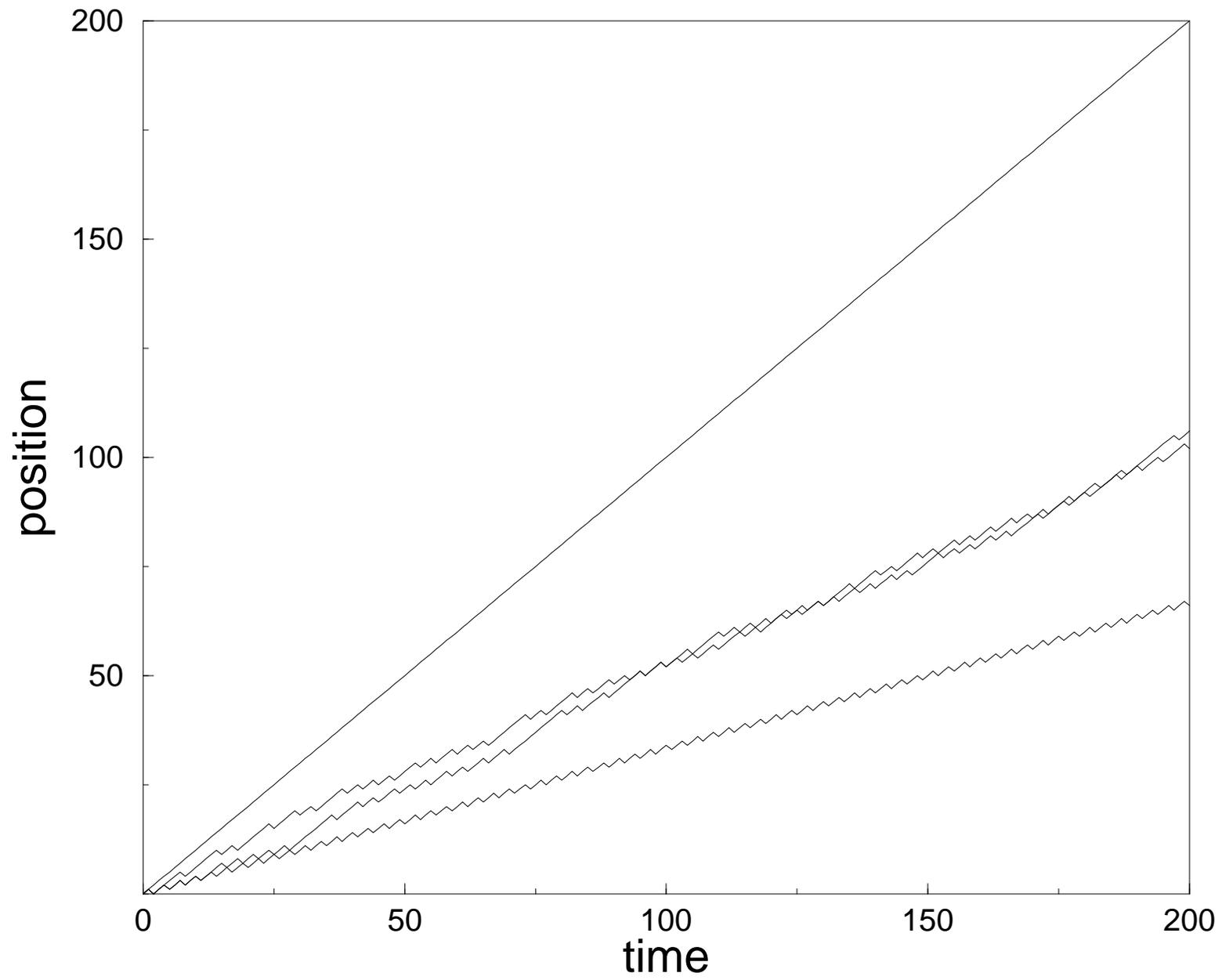
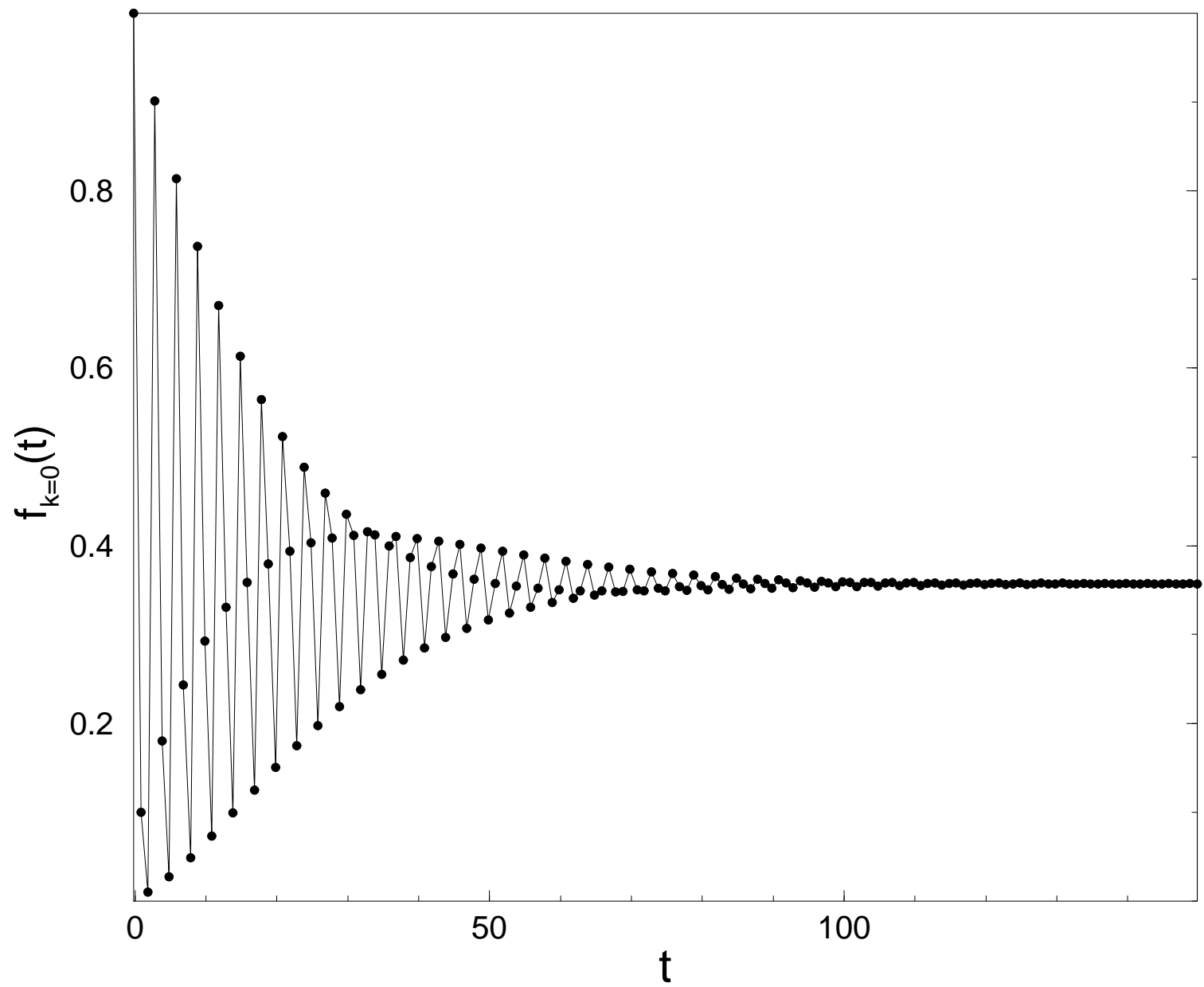
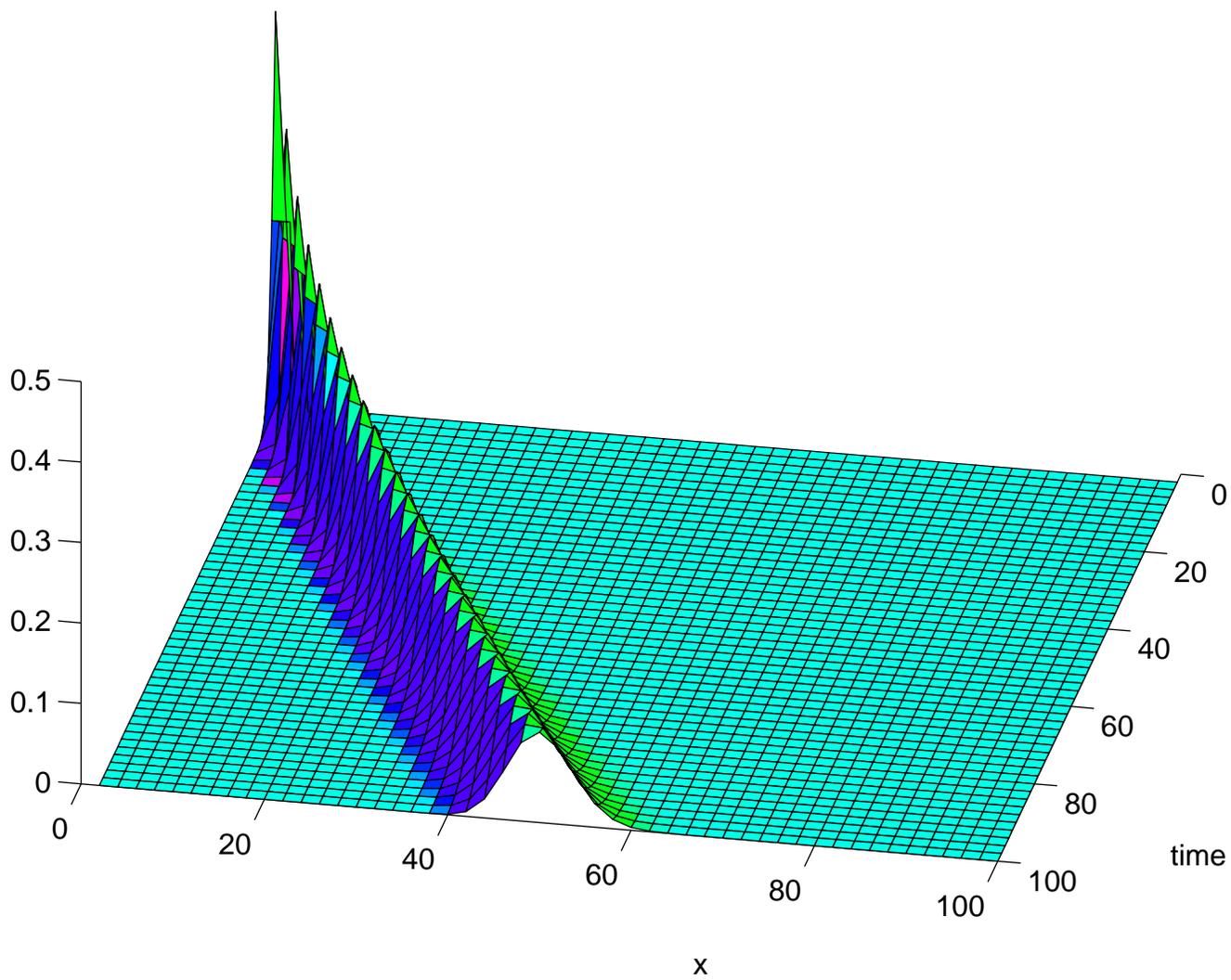
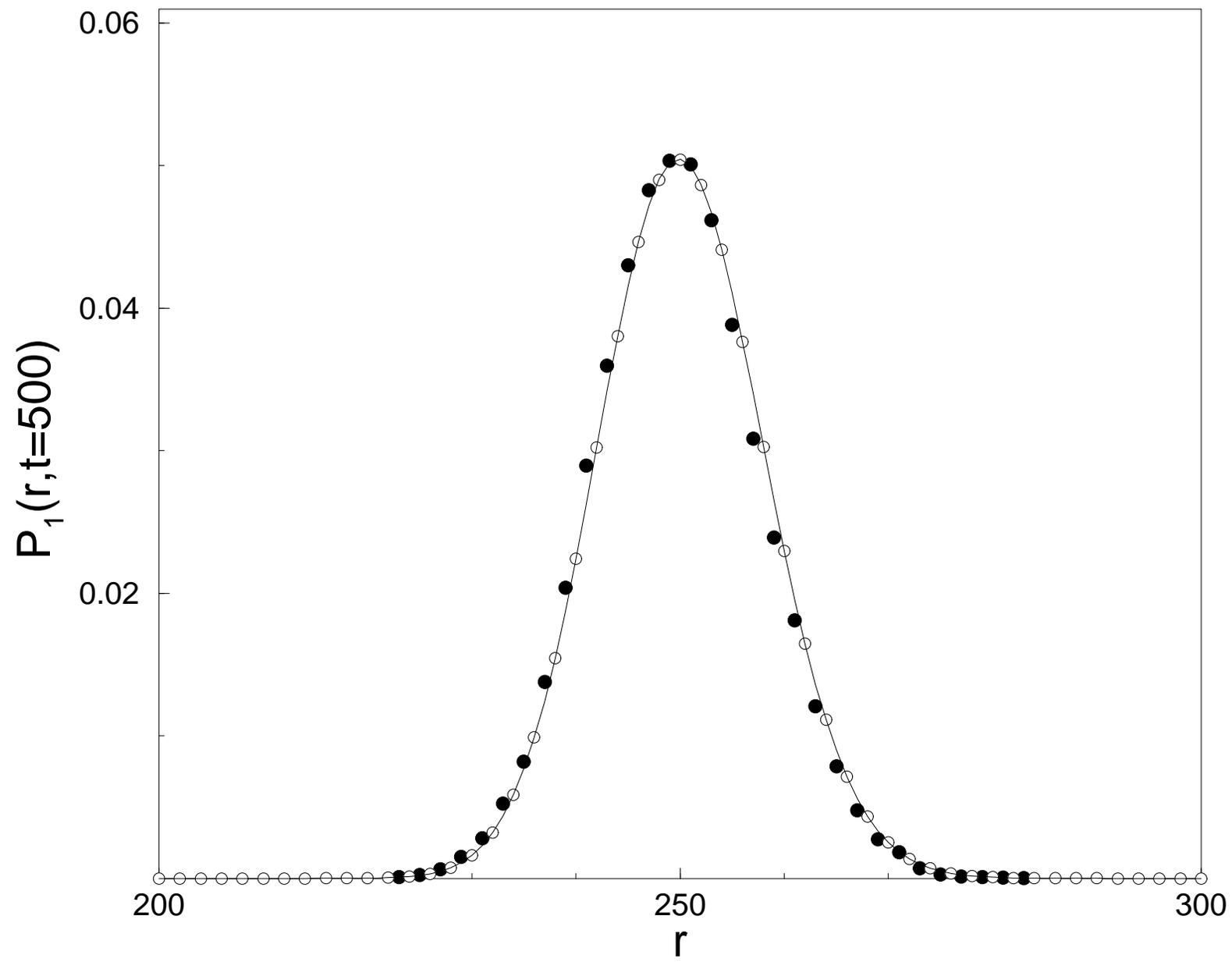
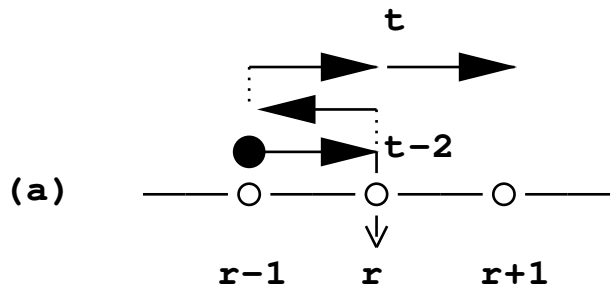
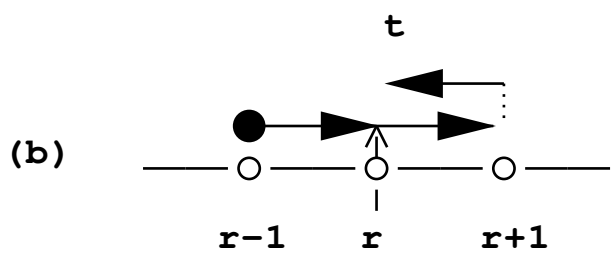
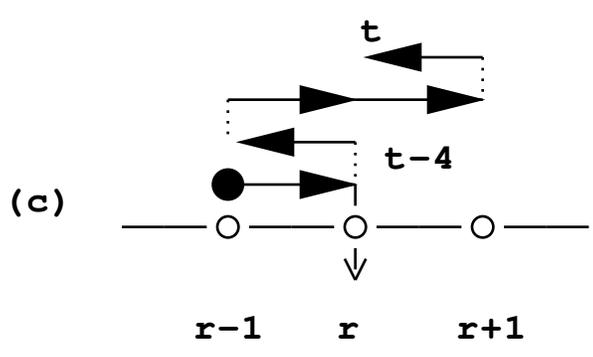
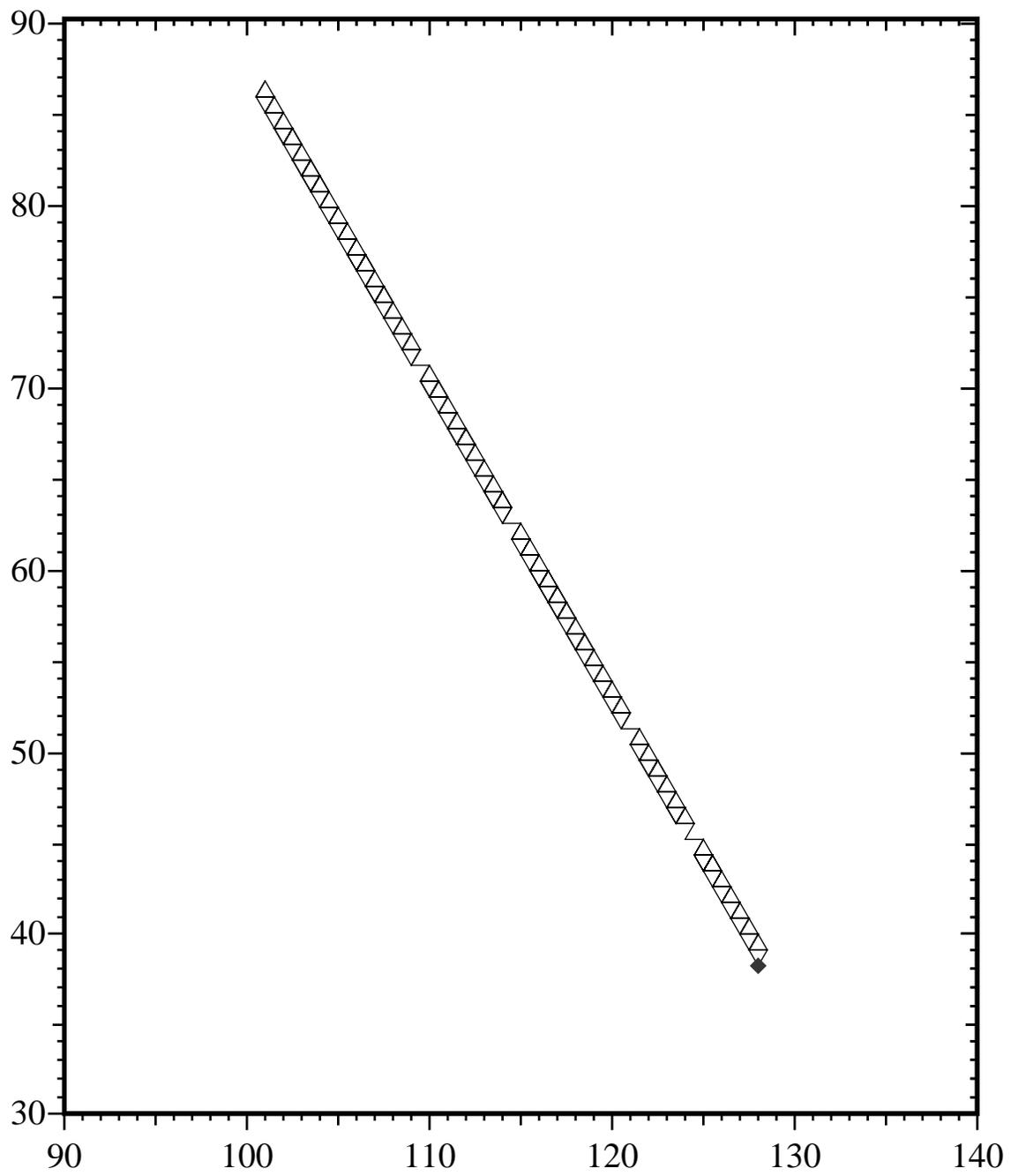
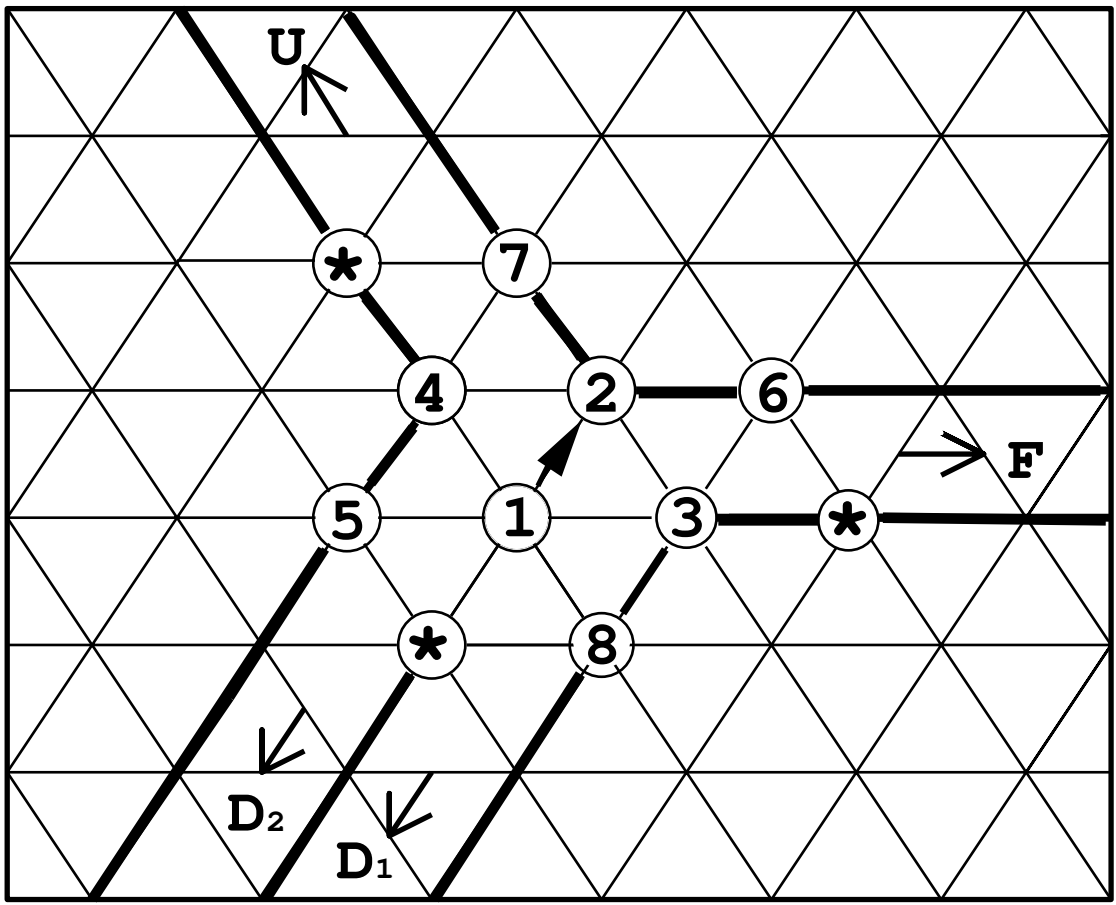
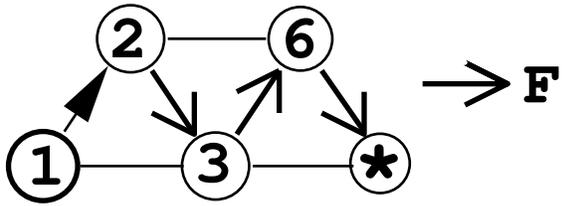
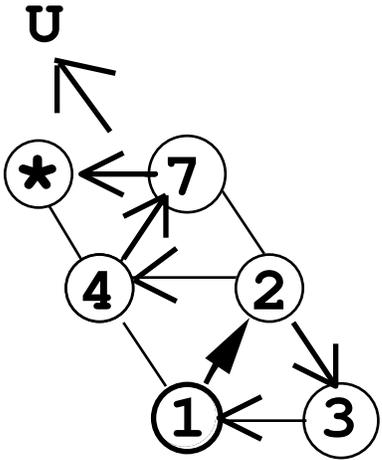
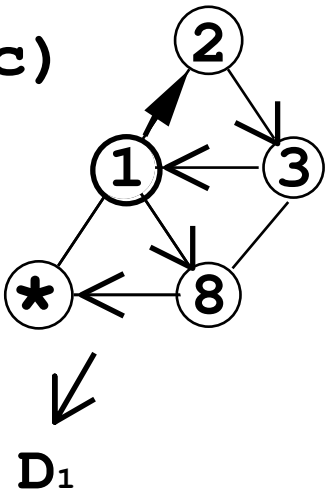
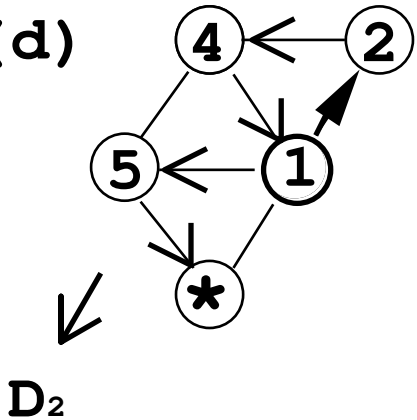
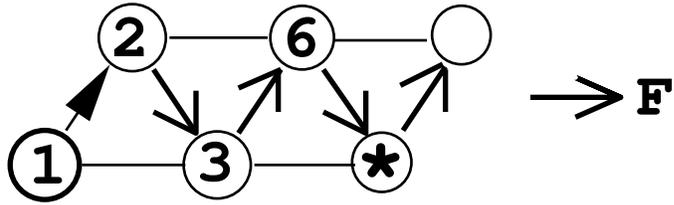
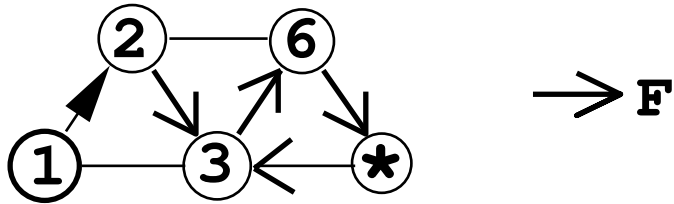
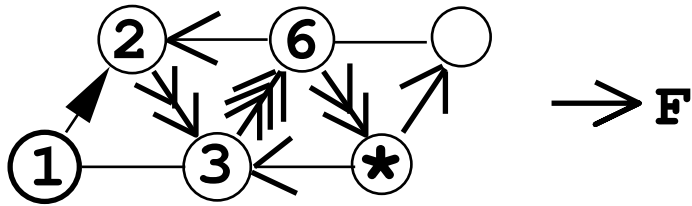
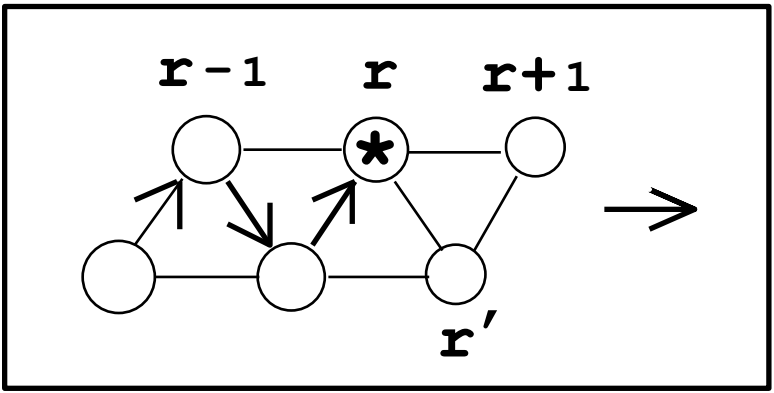
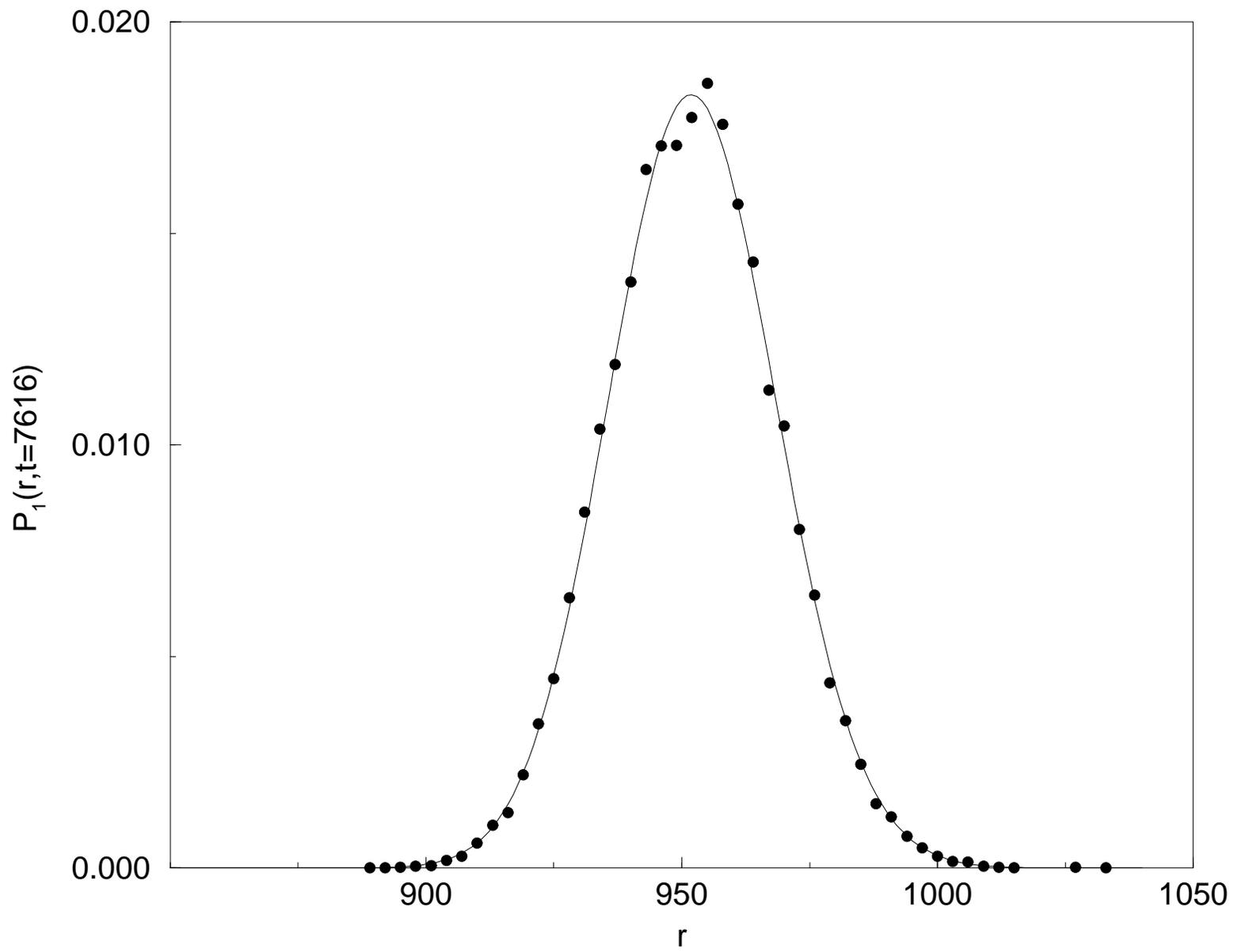
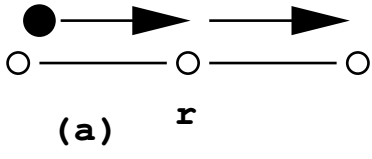
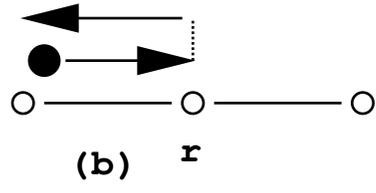
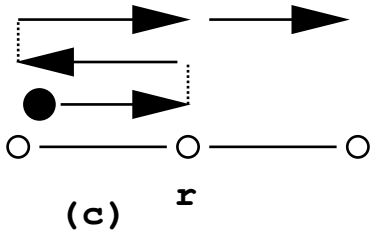
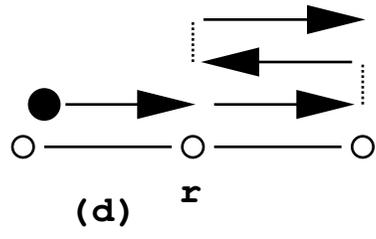
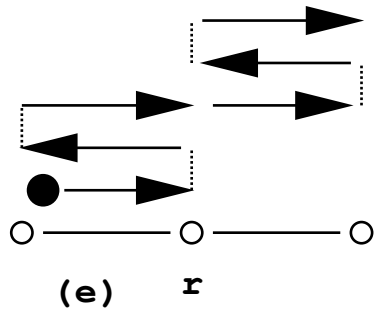
\end{document}